\documentclass[journal]{IEEEtran}
\usepackage{amssymb,amsfonts}
\usepackage{graphicx}
\usepackage{textcomp}
\usepackage{scalefnt}
\usepackage[cmex10]{amsmath}
\usepackage{subfig}
\usepackage{algorithm}
\usepackage{algpseudocode}
\usepackage[hyphens,spaces,obeyspaces]{url}
\usepackage{cleveref}
\usepackage{xcolor}

\newcommand{\diag}{\mathop{\rm diag}}
\newcommand{\tr}{\mathop{\rm tr}}

\newcommand{\Or}{\mathop{\mathcal{O}}}
\newcommand{\Sum}{\mathop{\rm sum}}
\newcommand{\zeros}{\mathop{\rm zeros}}
\newcommand{\nondiag}{\mathop{\rm nondiag}}
\newcommand{\desired}{\mathop{\rm desired}}
\newcommand{\signal}{\mathop{\rm signal}}
\newcommand{\intracell}{\mathop{\rm intracell}}
\newcommand{\intercell}{\mathop{\rm intercell}}
\newcommand{\interference}{\mathop{\rm interference}}

\graphicspath{{figures/}}

\ifCLASSINFOpdf

\else

\fi

\begin{document}

\title{\huge Joint AGC and Receiver Design for Large-Scale MU-MIMO Systems Using Low-Resolution Signal Processing in C-RANs}

\author{T. E. B. Cunha$^{1}$, R. C. de Lamare$^{1}$, T. N. Ferreira$^{2}$ and L. T. N. Landau$^{1}$ \\
$^{1}$Centre for Telecommunications Studies (CETUC), PUC-Rio, \\ Rio de Janeiro - 22451-900, Brazil \\
$^{2}$Engineering School, Fluminense Federal University (UFF), \\ Niterói, RJ - 24210-240, Brazil \\
Email: \{thiagoelias, delamare, lukas.landau\}@cetuc.puc-rio.br, tadeu\_ferreira@id.uff.br \vspace{-0.1em}
\thanks{The authors would like to thank the CAPES, CGI, CNPq, and FAPERJ
Brazilian agencies for funding.}
}


\maketitle

\begin{abstract}
Large-scale multi-user multiple-input multiple-output (MU-MIMO) systems and cloud
radio access networks (C-RANs) are considered promising technologies for the fifth generation (5G) of wireless networks. In these technologies, the use of low-resolution analog-to-digital converters (ADCs) is key for energy efficiency and for complying with constrained fronthaul links. Processing signals with few bits implies a significant performance loss and, therefore, techniques
that can compensate for quantization distortion are fundamental. In wireless systems, an automatic gain control (AGC) precedes the ADCs to adjust the input signal level in order to reduce the impact of quantization. In this work, we propose the joint optimization of the AGC, which works in the remote radio heads (RRHs), and a low-resolution aware (LRA) linear receive filter based on the minimum mean square error (MMSE), which works in the cloud unit (CU), for large-scale MU-MIMO systems with coarsely quantized signals. We develop linear and successive interference cancellation (SIC) receivers based on the proposed joint AGC and LRA MMSE (AGC-LRA-MMSE) approach. An analysis of the achievable sum rates along with a computational complexity study is also carried out. Simulations show that the proposed AGC-LRA-MMSE design provides substantial gains in bit error rates and achievable information rates over existing techniques.
\end{abstract}


\IEEEpeerreviewmaketitle

\section{Introduction}

In recent years, the widespread use of smartphones and
bandwidth-intensive applications and services has led to an exponential traffic growth in wireless networks \cite{cisco}. 5G has been developed to cope with
this growth and, at the same time, minimize the network capital and operating expenditures \cite{5G,c_ran_um}. In order to achieve a substantial increase in capacity, spectral and energy efficiencies, and in average cell throughput, solutions such as cloud radio access
networks (C-RANs) and large-scale MU-MIMO will be jointly deployed in the next generation systems
\cite{key_tec}.

In the traditional cellular network model, each base station (BS) covers a cell, receives, processes and transmits signals to and from the users \cite{c_ran_dois}. In the future, the huge number of devices connected to such networks will require the deployment of more BSs to meet the growing traffic demand. However, the deployment
of more BSs results in the increase of inter-cell interference and power consumption due to the BSs' hardware and cooling systems. In this context, C-RANs are a promising network architecture that performs centralized processing for next-generation systems
\cite{c_ran_tres, c_ran_cinco, C_RAN_12, pan2017, panjsac2017, c_ran_quatro,c_ran_zero}. In this
centralized architecture, BSs are broken down into low-cost Remote Radio Heads (RRHs) and a pool of Base Band Units (BBUs) located within a cloud unit (CU) \cite{c_ran_zero,pan2017}. The RRHs consist of simple radio antennas and active radio frequency components that perform transmit/receive signal processing such
as frequency conversion, power amplification and Analog-to-Digital (A/D) or Digital-to-Analog (D/A) conversion \cite{c_ran_tres}. The signal
processing tasks of each BS are migrated to the BBU pool, which is responsible for all the baseband signal processing \cite{c_ran_quatro}. Centralization aids network coordination and management, and can bring benefits such as reduction in the cost of
operating the network due to fewer site visits, energy consumption due to hardware and air-conditioning and easy upgrades \cite{c_ran_cinco}. C-RANs have received a great deal of attention in recent years thanks to their ability to improve the network performance with joint signal processing techniques that span
multiple base stations. Therefore, it mitigates the inter-cell interference in an efficient way, and in turn, allowing for higher spectral efficiency (SE)\cite{C_RAN_12}. However, one of the main
challenges to implement C-RANs is the limited capacity of fronthaul (FH)
links \cite{c_ran_zero,panjsac2017}.

Large-scale MU-MIMO systems can provide substantial gains over small-scale MU-MIMO systems in both energy and spectral efficiency \cite{ls_mimo}. In such systems a large number of antennas is employed at the BS to exploit the degrees of freedom and reduce the transmit power per antenna. However, the large number of antennas increases considerably the hardware cost and the power consumption due to the presence of A/D converters (ADCs) and D/A converters (DACs)
\cite{throughput_analysis,IQ_balance,mixed_adc}. The power
consumption in an uplink receiver design is heavily dependent on the
ADCs processing unit and the digital baseband processing unit, which
are both affected by the resolution in bits of the ADCs.
Specifically, the ADCs' power consumption scales linearly in the
sampling rate and exponentially in the number of bits
\cite{throughput_analysis}. Thus, we can reduce the power
consumption at the receiver using low-resolution ADCs. Furthermore,
the adoption of ADCs with fewer bits allows reduction in power
consumption, faster signal processing, cheaper systems, and
alleviates the capacity bottleneck of FH by reducing the
number of bits prior to transmission.

Quantizing signals with a low number of bits reduces the signal
quality due to the severe nonlinear distortion introduced.
In~\cite{adc_uniform_chosen_1,Modified,mixed_adc,roth} the
quantization process is shown to increase the MSE on the channel
estimation at the receiver. The quantization error can be
categorized into two types of distortions, the granular distortion,
and the clipping or overload distortion
\cite{quant_distort,asymptotic}. The granular distortion occurs when
the input signal lies within the quantizer-permitted range. The
overload distortion happens when the input signal exceeds the
allowed range, resulting in the clipping of the input signal. In
practice, ADCs are usually preceded by an AGC variable gain
amplifier, which aims to minimize the overload distortion
\cite{agc_system} or the effects of a too small input signal. The
AGC adjusts the analog signal level to the dynamic range of the
ADCs, which is important in scenarios where the received power
varies over time such as in mobile systems. In~\cite{b2} the effects
of choosing an adequate output of an AGC prior to quantization are
analyzed. Therefore, the AGC is an essential building block in the
receiver chain that implies an ADC because otherwise the ADC most
likely operates in a suboptimal operating point and, in particular,
the AGC design is key in large-scale MU-MIMO systems which employ
low-resolution ADCs.

The distortion produced by the quantization process and its impact
on the performance of communication systems has been studied in the
literature
\cite{throughput_analysis,IQ_balance,mixed_adc,quant_distort,asymptotic,agc_system,Modified,mixed_adc,roth,b2,b10,uplink_rates,Mixed_ADC_Massive,b3,ZShao,JMax,large_scale_4,adc_uniform_chosen_1,imperfet_CSI,imperfet_CSI_2,iswcs}.
Nevertheless, few studies address the design of AGCs. In
\cite{Modified} and \cite{roth}, modified MMSE receivers that take
into account the quantization effects in a MIMO system are presented
but they do not take into account the presence of an AGC. The
effects of a deterministic AGC on a quantized MIMO system with a
standard Zero-Forcing receive filter at the receiver were examined
in \cite{b2}. However, the work in \cite{b2} has not optimized the
AGC nor used a detector that considers the quantization effects.
Moreover, neither \cite{Modified,roth} nor \cite{b2} have been
designed for large-scale MU-MIMO with C-RANs. In
\cite{throughput_analysis} a suboptimal choice of the set of
quantization labels and thresholds was proposed with a rescheduling
scheme of the set of labels found through the \textit{Lloyd-Max}
algorithm. This analysis avoids the use of an AGC but the
\textit{Lloyd-Max} algorithm requires the probability density
function of the received signal to compute the optimum set of
labels, which is not practical. Therefore, novel techniques to deal
with the quantization effects for C-RANs are required.

In this work, we develop an uplink framework for jointly designing
the AGCs that work in the RRHs and low-resolution aware (LRA) linear
receive filters according to the MMSE criterion that works in the
CU. We then propose a joint AGC and LRA MMSE (AGC-LRA-MMSE) design
approach based on alternating optimization that adjusts the
parameters of the AGC and the receive filter. Based on the
AGC-LRA-MMSE aproach we then devise linear and SIC receivers. SIC is
a well-known layered detection scheme where a symbol is detected at
each layer \cite{v-blast,D_tse,sic_1,sic_2}. SIC detection improves
the detection accuracy and thus can help the receiver to achieve a
good performance even when dealing with coarsely quantized signals.
Unlike existing approaches with deterministic AGCs \cite{b2} and
modified MMSE receive filters \cite{b2,Modified,roth}, the proposed
AGC-LRA-MMSE approach jointly optimizes the AGCs and the receive
filters based on a statistical criterion using alternating
optimization. We report in \cite{iswcs} optimistic results of the
proposed joint optimization of the AGC and linear receivers in a
MU-MIMO scenario. However, \cite{iswcs} does not consider the SIC
detection scheme, the imperfect CSI, the imperfect knowledge about
the AGC coefficients at the CU, and does not evaluate the
performance of the proposed technique in a large-scale MU-MIMO
system in C-RANs that are considered by this paper. We also derive
expressions for the achievable sum rates and evaluate the
computational complexity of the proposed AGC-LRA-MMSE approach.
Simulations show that the proposed AGC-LRA-MMSE design provides
substantially better error rates and higher achievable rates than
existing techniques. The main contributions of this work can be
summarized as:
\begin{enumerate}
\item{An uplink framework for jointly designing MMSE-based AGCs that work in
the RRHs and LRA linear receivers.}

\item{Linear and SIC receivers based on the proposed AGC-LRA-MMSE approach.}

\item{Analytical expressions for the achievable sum rates and the
computational complexity of the proposed AGC-LRA-MMSE approach.}
\end{enumerate}

The organization of this paper is as follows. The next section
details the system model of the large-scale MU-MIMO system with the C-RAN signal processing scheme, states the problem and describes the properties of the quantizer adopted and a model for the distortion produced by the quantization process. The proposed AGC-LRA-MMSE design approach is presented
along with the details of its derivation and computational
complexity in Section III. Section IV develops the sum rate analysis
of the proposed AGC-LRA-MMSE approach. Simulation results are
presented and discussed in Section V. The conclusions of this paper
are given in Section VI.

\textit{Notation:} Vectors and matrices are denoted by lower and
upper case italic bold letters. The operators $(\cdot)^T$,
$(\cdot)^H$ and $\tr(\cdot)$ stand for transpose, Hermitian
transpose and trace of a matrix, respectively. $\mathbf{1}$ denotes
a column vector of ones and $\mathbf{I}$ denotes an identity matrix.
The operator $\mathrm{E}[\cdot]$ stands for expectation with respect to the
random variables and the operator $\odot$ corresponds to the
Hadamard product. Finally, $\diag(\mathbf{A})$ denotes a diagonal
matrix containing only the diagonal elements of $\mathbf{A}$ and
$\nondiag(\mathbf{A})=\mathbf{A}-\diag(\mathbf{A})$. The operators
$\mbox{Q}(\cdot)$ and $\mathrm{DEC}(\cdot)$ represent respectively the
quantization of a vector with an arbitrary number of bits and the
slicer used for detection.
\section{System Description}
\begin{figure}[!h]
\centering
\includegraphics[scale=0.45]{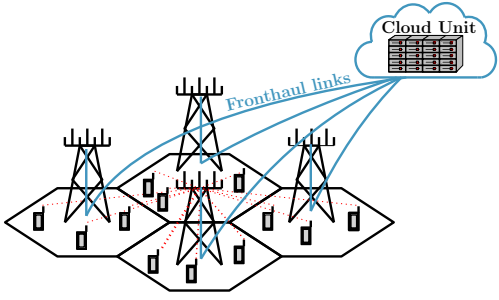}
\caption{Large-scale MU-MIMO uplink system with centralized signal processing.}
\label{Fig:1}
\end{figure}
Let us consider the uplink of a large-scale MU-MIMO system with
C-RANs. As shown in Fig. \ref{Fig:1}, the system consists of one BBU
pool located at the CU and one cluster with $L$ cells in which each
cell has one RRH at the center of the cell and a total of $K$ users
randomly distributed in their covered area. It is assumed that each
RRH is equipped with $N_R$ receive antennas, while each user is
equipped with $N_T$ transmit antennas. It is further assumed that
each RRH is connected to the CU via an imperfect finite-capacity
digital FH link.

In the uplink, all users simultaneously communicate with the RRHs in
the same time-frequency resource. After that, each RRH processes the
signal received from all users with independent AGCs, quantizes the
AGC output signal and then forwards the quantized signal and also
AGC coefficients to the CU via their FH link. {Since FH links have
high capacity and the number of coefficients is equal to the number
of antennas times the number of bits per coefficient, then  the load
can be considered modest.} At last, the BBU pool jointly detects the
users' messages based on signals from all RRHs. The steps of the
proposed joint AGC and LRA-MMSE detection scheme are summarized in
Algorithm \ref{alg:proposed_scheme}. \vspace{-8pt}
\begin{algorithm}[H]
\caption{AGC and LRA-MMSE scheme}
\begin{algorithmic}[1]
\State Each RRH receives the analog signals transmitted by all users of the cluster.
\State The $l$-th RRH computes its optimal AGC coefficients.
\State The analog signal received at the $l$-th RRH is sampled, processed by the AGC and then is quantized with few bits of resolution.
\State The quantized signals and the AGC coefficients are sent to the CU through imperfect FH links.
\State Due to the transmission errors in the FH links, the optimal AGC coefficients are assumed to be distorted.
\State The LRA-MMSE receive filter is computed at the CU taking into account the quantization distortion and the presence of the AGCs at the RRHs.
\State The symbols of the users are estimated at the CU by linear or SIC detection schemes.
\end{algorithmic}
\label{alg:proposed_scheme}
\end{algorithm}
\vspace{-10pt} As the users transmit their signals in the same
time-frequency resource, such signals can interfere with each other
resulting in the intracell interference and in the intercell
interference. In order to model this scenario the $N_R$-dimensional
received signal vector at the $l$-th  RRH can be expressed as
\begin{eqnarray}
\mathbf{y}_{l}=\underbrace{\mathbf{H}_{ll}^{(k)}\mathbf{x}_l^{(k)}}_{\stackrel{\desired}{\signal}} + \underbrace{\sum_{\stackrel{u=1}{u\neq k}}^{K}\mathbf{H}_{ll}^{(u)}\mathbf{x}_l^{(u)}}_{\stackrel{\intracell}{\interference}}
+\underbrace{\sum_{\stackrel{i=1}{i\neq l}}^{L}\sum_{u=1}^{K}\mathbf{H}_{li}^{(u)}\mathbf{x}_i^{(u)}}_{\stackrel{\intercell}{\interference}}+\mathbf{n}_{l},
\label{eq:1}
\end{eqnarray}
\noindent where $\mathbf{H}_{li}^{(u)} \in \mathbb{C}^{N_R
\times N_T}$ is the channel matrix between the $N_T$ transmit
antennas of the $u$-th user in the $i$-th cell and the $N_R$ receive
antennas of the $l$-th RRH; $\mathbf{x}_i^{(u)} \in \mathbb{C}^{N_T \times 1}$ is the transmitted vector by the $u$-th user in the $i$-th cell; and $\mathbf{n}_l \in
\mathbb{C}^{N_R \times 1}$ contains additive
white Gaussian noise (AWGN) at the $l$-th RRH. The
elements of both $\mathbf{x}_i^{(u)}$ and $\mathbf{n}_l$ are assumed to be independent and identically distributed (i.i.d) zero mean circularly
symmetric complex Gaussian (ZMCSCG) random variables with variances equal to unit and $\sigma_n^2$, respectively.

The channel matrix $\mathbf{H}_{li}^{(u)}$ models independent fast
fading, geometric attenuation, and log-normal shadow fading
\cite{large_scale_4,large_scale_3,large_scale_1,large_scale_2}.
Under perfect channel state information at the receiver (CSIR)
assumption, the coefficients of the channel matrix are given by
$\mathbf{H}_{li}^{(u)}=\tilde{\mathbf{H}}_{li}^{(u)}\sqrt{\beta_{li}^{(u)}}$,
where each entry $\tilde{h}_{limn}^{(u)}$ of
$\tilde{\mathbf{H}}_{li}^{(u)}$ represents the fast fading
coefficient between the $n$-th transmit antenna of the $u$-th user
in the $i$-th cell to the $m$-th receive antenna of the $l$-th RRH.
Its coefficients are assumed to be i.i.d ZMCSCG random variables
with unit variance. The quantity $\beta_{li}^{(u)}$ represents the
geometric attenuation and shadow fading, which is assumed to be
independent over $m$ and $n$. This factor is modeled as follows
\begin{eqnarray}
\beta_{li}^{(u)}=\frac{z_{li}^{(u)}}{\left(d_{li}^{(u)}\right)^\gamma},
\label{eq:large_scale}
\end{eqnarray}
\noindent where $z_{li}^{(u)}$ represents the shadow fading and
obeys a log-normal distribution with standard deviation
$\sigma_{\text{shadow}}$ (i.e, $10\log\left(z_{li}^{(u)}\right)$
follows a Gaussian distribution with zero-mean and standard
deviation $\sigma_{\text{shadow}}$), $d_{li}^{(u)}$ corresponds to
the distance between the $u$-th user in the $i$-th cell and the RRH in
the $l$-th cell measured in kilometers, and $\gamma$ is the
path-loss exponent.

In this work we consider a time block fading model where the
small-scale channel matrix $\tilde{\mathbf{H}}_{li}^{(u)}$ stays
constant during the \textit{coherence interval} of a data packet,
and the large-scale coefficient $\beta_{li}^{(u)}$ stays constant
during a \textit{large scale coherence interval} of a block of data
packets \cite{coherence}. A usual assumption is that one block of
data packets corresponds to a set of 40 data packets, i.e, 40 sets
of symbols of a given modulation scheme. The matrices
$\tilde{\mathbf{H}}_{li}^{(u)}$ and the coefficients
$\beta_{li}^{(u)}$ are assumed to be independent in different
coherence intervals and large scale coherence intervals,
respectively.

In practice, CSI is obtained through channel estimation and thus is
inevitably contaminated by noise. The imperfect CSI at the RRH can
be modeled through a Gauss-Markov uncertainty of the form
\cite{imperfet_CSI,imperfet_CSI_2}
\begin{eqnarray}
\hat{\mathbf{H}}_{\mathrm{est}}=\sqrt{1-\eta^2}\tilde{\mathbf{H}}+\eta\mathbf{E},
\label{eq:erros_estimacao_canal}
\end{eqnarray}
\noindent
where $\hat{\mathbf{H}}_{\mathrm{est}}$ represents the imperfect estimation of $\tilde{\mathbf{H}}$ and $\mathbf{E}\sim \mathcal{CN}(\mathbf{0},\mathbf{I})$ corresponds to the channel estimation errors modeled as an AWGN. The CSI related parameter $\eta$ characterizes the channel estimation errors. Specifically, $\eta=0$ means perfect channel knowledge, the values of $0<\eta<1$ correspond to partial CSI and $\eta=1$ accounts for no channel knowledge. The large-scale fading can be accurately estimated due to their much slower variation in comparison to the symbol rate.  Therefore, we assume that the coefficients $\beta_{li}^{(u)}$ are known.
{Imperfect CSI is then considered and their impact will be evaluated through the channel estimation errors model. We remark that, in addition to the imperfect CSI due to channel estimation techniques and their associated errors, there is an impact of the low-resolution ADCs on the channel estimation. Since our work is focused on the joint AGC and receiver design, we have opted to account for imperfect CSI due to both channel estimation and low-resolution ADCs with an error model based on Gaussian random variables. We advocate that this is reasonable for large-scale systems due to the central limit theorem by which an error model based on Gaussian random variables is sufficient to describe the impact of errors related to both estimation methods and low-resolution ADCs. Exploring channel estimation techniques for the considered scenario and the proposed AGC-LRA-MMSE approach deserves a thorough study and development effort, and thus is left as future work.} 

The received signal at the $l$-th RRH can be rewritten as
\begin{eqnarray}
\mathbf{y}_{l} &=&\sum_{i=1}^{L}\sum_{u=1}^{K}\mathbf{H}_{li}^{(u)}\mathbf{x}_i^{(u)}+\mathbf{n}_{l} \notag \\
&=& {\mathbf{H}}_{l}\mathbf{x}+\mathbf{n}_l,
\label{eq:received_l_matrix}
\end{eqnarray}
\noindent where
$\hat{\mathbf{H}}_l=[\hat{\mathbf{H}}_{l1},...,\hat{\mathbf{H}}_{lL}]$
is the $\mathbb{C}^{N_R \times LKN_T}$ matrix with the coefficients
of the channels between all users of the cluster and the $N_R$
receive antennas of the $l$-th RRH; and
$\mathbf{x}=[\mathbf{x}_1^T,...,\mathbf{x}_{L}^T]^T$
is the $\mathbb{C}^{LKN_T \times 1}$ transmit vector by all
users of the cluster.

In the considered scenario each RRH forwards their signals to the CU
through digital FH links with $R_v=b$ bits, $\forall v \in
\{1,..,V\}$, corresponding to the quantization level $V=2^b$. For a
complex-value signal $y_{i} \in \mathbb{C}$, $1 \leq i \leq N_R$, we
quantize the real and the imaginary part separately. The
quantization operation $\mathrm{Q}(y_{i,j})$, $j \in
\{\mathrm{R},\mathrm{I}\}$, of the real or imaginary parts of the
input signal is a nonlinear mapping of $y_{i,j} \in \mathbb{R}$ to a
discrete set that results in additive distortion that follows the
given rule \vspace{-8pt}
\begin{eqnarray}
r_{i,j}=\mathrm{Q}(y_{i,j})= y_{i,j}+q_{i,j}.
\end{eqnarray}
\noindent where $q_{i,j}$ corresponds to the resulting quantization error.

In this paper we consider the use of uniform symmetric mid-riser
quantizers characterized by a set of $V+1$ input thresholds
$\mathcal{T}_b=\{\tau_1,\tau_2,...,\tau_{V+1}\}$, and a set of $V$
output labels $\mathcal{A}=\{a_1,a_2,...,a_{V}\}$ where $a_v \in
(\tau_v,\tau_{v+1}]$ \cite{asymptotic}. Uniform quantization has
been chosen because it allows for simple and tractable modeling and
is probably the most widely used quantization technique in practice
\cite{agc_system,adc_uniform_chosen_1}. The output levels of the
quantizer are assigned as
$a_v=\frac{-V\Delta}{2}+(v-\frac{1}{2})\Delta$, where $\Delta$ is
the quantizer step-size. The input thresholds are given by
$\tau_1=-\infty$, $\tau_{V+1}=\infty$, and
$\tau_v=a_v+\frac{\Delta}{2}$, for $v=2,3,...,V$.

The quantization factor $\rho_q^{(i,j)}$ indicates the relative
amount of granular noise that is generated when the analog signal is
quantized. This factor was defined in \cite{Modified} as follows
\begin{eqnarray}
\rho_q^{(i,j)}=\frac{\mathrm{E}[q^2_{i,j}]}{r_{y_{i,j}y_{i,j}}},
\end{eqnarray}
\noindent where $r_{y_{i,j}y_{i,j}}=\mathrm{E}[y_{i,j}^2]$ is the variance of
$y_{i,j}$. This factor depends on the number of quantization bits
$b$, the quantizer type and the probability density function of
$y_{i,j}$. Here, the uniform quantizer design is based on minimizing
the MSE between the input $y_{i,j}$ and the output $r_{i,j}$. Under
the optimal design of the scalar finite resolution quantizer, the
following equations hold for all $1 \leq i \leq N_R$, $j \in
\{\mathrm{R},\mathrm{I}\}$,\cite{Modified}:
    \begin{eqnarray}
    \mathrm{E}[q_{i,j}]&=&0, \\
    \mathrm{E}[r_{i,j}q_{i,j}]&=&0,\\
    \mathrm{E}[y_{i,j}q_{i,j}]&=&-\rho_q^{(i,j)}r_{y_{i,j}y_{i,j}}.
    \end{eqnarray}

    Under multipath propagation conditions and for a large number of
antennas $y_{i,j}$ are approximately Gaussian distributed and  they
undergo nearly the same distortion factor $\rho_{i,j}=\rho_q$,
for all $i$ and for all $j$. In this work the scalar uniform quantizer
processes the real and imaginary parts of the input signal $y_{i,j}$
in a range $\pm \frac{\sqrt{b}}{2}$. Let $q_i=q_{i,\mathrm{R}}+jq_{i,\mathrm{I}}$ be
the complex quantization error and under the assumption of
uncorrelated real and imaginary parts of $y_i$ we obtain
\begin{eqnarray}
    r_{q_{i}q_{i}}&=&\mathrm{E}[q_{i}q_{i}^*]=\rho_q r_{y_{i}y_{i}}, \nonumber \\
    r_{y_{i}q_{i}}&=&\mathrm{E}[y_{i}q_{i}^*]=-\rho_q r_{y_{i}y_{i}}.
    \label{eq:181}
\end{eqnarray}

    As shown in \cite{asymptotic}, the optimal quantization step
$\Delta$ for the uniform midriser quantizer case and for a Gaussian
source decreases as $\sqrt{b}2^{-b}$ and $\rho_q$ is asymptotically
well approximated by $\frac{\Delta^2}{12}$.

\section{Joint AGC and LRA-MMSE Receive Filter Design}

\begin{figure}[!h]
   \centering
    \includegraphics[scale=0.55]{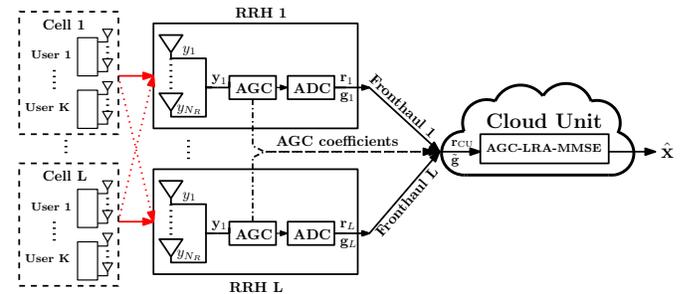}
    \caption{Large-scale MIMO system with C-RAN and AGC.}
    \label{Fig:AGC_CRAN_scheme}
    \end{figure}

The joint AGC and LRA-MMSE receive filter design consists of an
alternating optimization based on the MMSE criterion that jointly
computes the AGC matrices that work at each RRH and the LRA-MMSE
receive filter that works at the CU \cite{niesen,jidf}. The AGCs are
used before the quantizers to reduce the distortion arising from the
low-resolution quantization. {After the quantized signals and the
AGC coefficients are sent to the CU via the FH links, the desired
symbols are estimated by an LRA-MMSE receive filter, which allows
the use of numerous detection, precoding and estimation techniques
\cite{jio,jidf,jiols,jiomimo},
\cite{smce,1bitce,TongW,jpais_iet,armo,badstc,baplnc,goldstein,qian,jio,jidf,jiols,jiomimo,dce,jpba},
\cite{mmimo,wence,deLamare2003,itic,deLamare2008,cai2009,jidf,jiomimo,Li2011,
wlmwf,dfcc,deLamare2013,did,rrmser,bfidd,1bitidd,aaidd,aalidd},
\cite{mmd,bamax,mwc_wsa,mwc,tds_cl,tds,armo,badstbc},
\cite{lclattice,switch_int,switch_mc,gbd,wlbd,mbthp,rmbthp,bbprec,1bitcpm,bdrs,baplnc,memd,wljio,locsme,okspme,lrcc}.
The FH links convey $N_R \times b$ bits to the CU, where $N_R$ is
the number of receive antennas at the RRHs and $b$ is the number of
bits. Therefore, a drawback of the proposed scheme is a small
increase in the traffic on the FH links. An example of a FH with
high capacity is reported in \cite{Song2015}, whereas a study with
imperfect FH links is described in \cite{Bartelt2017}.} The proposed
scheme is illustrated in Fig. \ref{Fig:AGC_CRAN_scheme}.
\vspace{-8pt}

We initialize the AGC-LRA-MMSE design by computing, for each RRH, the LRA-MMSE receiver that takes into account quantization and
a standard AGC as an identity matrix. In order to obtain the optimal AGC
coefficients, the derivative of the cost function with respect to the AGC coefficients that takes into account both the presence of an AGC and the LRA-MMSE receive filter
is computed in the following. At last, we derive the LRA-MMSE receive
filter which works in the CU and considers both the quantization effects and the presence of the AGCs. It is important to mention that no convergence problems
were observed in this alternating optimization, which results in accurate symbol estimates. This is quite intuitive since increasingly better estimates $\hat{\mathbf{x}}$ are
generated which gives rise to better estimates of the covariance matrices.
\subsection{Low-Resolution Aware Receive Filter (LRA-MMSE)}

The quantized signal vector at $l$-th RRH can be
expressed with the help of the Bussgang decomposition
\cite{bussgang} as the linear model
\begin{equation}
\begin{split}
\mathbf{r}_l & = \mathbf{y}_l+\mathbf{q}_l \\
& = \mathbf{G}_l\mathbf{y}_l+\mathbf{q}_l,
\end{split}
\end{equation}
where $\mathbf{G}_l \in \mathbb{R}^{N_R \times N_R}$ is a diagonal matrix with the real coefficients of the AGC and $\mathbf{q}_l \in
\mathbb{C}^{N_R \times 1}$ is the quantization noise vector. Note that the AGC matrix is assumed to be initialized as an identity matrix. Then, the linear receive filter $\mathbf{W}_l  \in \mathbb{C}^{KN_T \times
N_R}$ that minimizes the MSE
\begin{eqnarray}
\varepsilon
=\mathrm{E}[||\mathbf{x}_l-\hat{\mathbf{x}}_l||^2_2]=\mathrm{E}[||\mathbf{x}_l-\mathbf{W}_l\mathbf{r}_l||^2_2],
\label{mse_cost1}
\end{eqnarray}
\noindent is given by
\begin{eqnarray}
\mathbf{W}_l=\mathbf{R}_{x_lr_l}\mathbf{R}_{r_lr_l}^{-1}.
\label{eq:wiener}
\end{eqnarray}
\noindent where the cross-correlation matrix $\mathbf{R}_{x_lr_l} \in \mathbb{C}^{KN_T \times N_R}$, and the autocorrelation matrix $\mathbf{R}_{r_lr_l} \in \mathbb{C}^{N_R \times N_R}$ are expressed as
\begin{eqnarray}
\mathbf{R}_{x_lr_l}&=&\mathbf{R}_{x_ly_l}+\mathbf{R}_{x_lq_l},
\label{eq:Rxr} \\
\mathbf{R}_{r_lr_l}&=&\mathbf{R}_{y_ly_l}+\mathbf{R}_{y_lq_l}+\mathbf{R}_{y_lq_l}^H+\mathbf{R}_{q_lq_l}.
\label{eq:Rrr}
\end{eqnarray}

    We obtain the autocorrelation matrix $\mathbf{R}_{y_ly_l}$ and
the cross-correlation matrix $\mathbf{R}_{xy_l}$ with the
model presented by (\ref{eq:received_l_matrix}) as
\begin{eqnarray}
\mathbf{R}_{x_ly_l}&=&\mathrm{E}[\mathbf{x}_l\mathbf{y}_l^H]=\mathbf{R}_{x_lx_l}\hat{\mathbf{H}}_{ll}^H.  \label{eq:Rxy}\\
\mathbf{R}_{y_ly_l}&=&\mathrm{E}[\mathbf{y}_l\mathbf{y}_l^H]=\hat{\mathbf{H}}_l\mathbf{R}_{xx}\hat{\mathbf{H}}_l^H+\mathbf{R}_{n_ln_l}, \label{eq:Ryy}
\end{eqnarray}

    To compute (\ref{eq:Rxr}) and (\ref{eq:Rrr}) we need to obtain the
matrices $\mathbf{R}_{x_lq_l}$, $\mathbf{R}_{y_lq_l}$ and
$\mathbf{R}_{q_lq_l}$ as a function of the channel parameters and
the distortion factor $\rho_q$. Following the procedure developed in
\cite{Modified} the expressions of $\mathbf{R}_{x_lq_l}$,
$\mathbf{R}_{y_lq_l}$ and $\mathbf{R}_{q_lq_l}$ are given by
\begin{eqnarray}
\mathbf{R}_{x_lq_l}&=& -\rho_q\mathbf{R}_{x_ly_l},  \label{eq:Rxq} \\
\mathbf{R}_{y_lq_l} &\approx& -\rho_q\mathbf{R}_{y_ly_l}, \label{eq:Ryq}  \\
\mathbf{R}_{q_lq_l}&=& \rho_q \mathbf{R}_{y_ly_l}-(1-\rho_q)\rho_q \nondiag(\mathbf{R}_{y_ly_l}).\label{eq:Rqq}
\end{eqnarray}
where nondiag takes its input argument and sets its main diagonal elements to zero, as detailed in the Notation part of the Introduction.

    Substituting  (\ref{eq:Rxy}) and (\ref{eq:Rxq}) in (\ref{eq:Rxr}), and  (\ref{eq:Ryy}), (\ref{eq:Ryq}) and (\ref{eq:Rqq}) in (\ref{eq:Rrr}) we get
\begin{eqnarray}
\mathbf{R}_{x_lr_l}&\approx&(1-\rho_q)\mathbf{R}_{x_ly_l}. \label{eq:Rxr_2} \\
\mathbf{R}_{r_lr_l}&\approx&(1-\rho_q)(\mathbf{R}_{y_ly_l}-\rho_q\nondiag(\mathbf{R}_{y_ly_l})). \label{eq:Rrr_2}
\end{eqnarray}

    We can then apply (\ref{eq:Rxr_2}) and (\ref{eq:Rrr_2}) in (\ref{eq:wiener}) to obtain the expression of the LRA-MMSE receive filter at the $l$-th RRH
    \begin{eqnarray}
    \mathbf{W}_l \approx \mathbf{R}_{x_ly_l}(\mathbf{R}_{y_ly_l}-\rho_q\nondiag(\mathbf{R}_{y_ly_l}))^{-1}.
    \label{eq:l_rec_cell}
    \end{eqnarray}
\subsection{AGC Design}
    In the following the optimum AGC matrix $\mathbf{G}_l$ at the $l$-th RRH is computed by
minimizing the MSE. We consider $\mathbf{G}_l$ as a diagonal matrix with real
coefficients and $\mathbf{g}_l$ a vector with the diagonal
coefficients of $\mathbf{G}_l$. Therefore, we can write
$\mathbf{G}_l=\diag(\mathbf{g}_l)$. Since $\mathbf{G}_l$ is a
diagonal matrix with real coefficients we have
$\diag(\mathbf{g}_l)^H=\diag(\mathbf{g}_l)$. Then, the MSE cost
function that considers both the LRA-MMSE receive
filter and the AGC matrix at the $l$-th RRH can be rewritten as
\begin{eqnarray}
\varepsilon =\mathrm{E}[||\mathbf{x}_l-\mathbf{W}_l(\alpha\diag(\mathbf{g}_l)\mathbf{y}_l+\mathbf{q}_l)||^2_2],
\label{eq:cost_agc}
\end{eqnarray}
\noindent where $\alpha$ corresponds to the clipping factor of the
AGC. This factor is a commonly used rule to adjust the amplitude of
the received signal in order to minimize the overload distortion
\cite{quant_distort}. To solve this problem we compute the
derivative of (\ref{eq:cost_agc}) with respect to $\diag(\mathbf{g}_l)$
while keeping $\mathbf{W}_l$ fixed. Therefore, we will have an
initialization using the linear filter previously computed by
(\ref{eq:l_rec_cell}) in order to obtain $\mathbf{G}_l$. To
compute the optimum $\mathbf{G}_l$ matrix we compute the derivative
of (\ref{eq:cost_agc}) with respect to $\diag(\mathbf{g}_l)$, equate
the derivative terms to zero and solve it for $\mathbf{g}_l$:
    \begin{eqnarray}
    \frac{\partial \varepsilon}{\partial \mathbf{g}_l}&=&-\alpha\underbrace{\frac{\partial \tr(\mathbf{R}_{x_ly_l}\diag(\mathbf{g}_l)\mathbf{W}_l^H)}{\partial \mathbf{g}_l}}_{\mathrm{I}} \nonumber \\
    &&-\alpha\underbrace{\frac{\partial\tr(\mathbf{W}_l\diag(\mathbf{g}_l)\mathbf{R}_{x_ly_l}^H)}{\partial \mathbf{g}_l}  }_{\mathrm{II}} \nonumber \\
           &&+\alpha^2\underbrace{\frac{\partial \tr(\mathbf{W}_l\diag(\mathbf{g}_l)\mathbf{R}_{y_ly_l}\diag(\mathbf{g}_l)\mathbf{W}_l^H)}{\partial \mathbf{g}_l}}_{\mathrm{III}}\nonumber \\
           &&+\alpha\underbrace{\frac{\partial \tr(\mathbf{W}_l\diag(\mathbf{g}_l)\mathbf{R}_{y_lq_l}\mathbf{W}_l^H)}{\partial \mathbf{g}_l}}_{\mathrm{IV}} \nonumber \\
           &&+\alpha\underbrace{\frac{\partial \tr(\mathbf{W}_l\mathbf{R}_{y_lq_l}^H\diag(\mathbf{g}_l)\mathbf{W}_l^H)}{\partial \mathbf{g}_l}}_{\mathrm{V}}.  \label{eq:deriv_1}
    \end{eqnarray}

    The details of the derivation are presented in the Appendix. The
results are given by
    \begin{align}
    \mathrm{I} &= [(\mathbf{R}_{x_ly_l}^T\odot\mathbf{W}_l^H)\mathbf{1}], \\
    \mathrm{II} &= [(\mathbf{R}_{x_ly_l}^H\odot\mathbf{W}_l^T)\mathbf{1}],
    \end{align}
    \begin{align}
    \mathrm{III}&= [(\mathbf{W}_l^T\odot(\mathbf{R}_{y_ly_l}\diag(\mathbf{g}_l)\mathbf{W}_l^H))\mathbf{1}] \nonumber \\
    &\hspace{12pt}+[((\mathbf{R}_{y_ly_l}^T\diag(\mathbf{g}_l)\mathbf{W}_l^T)\odot\mathbf{W}_l^H)\mathbf{1}], \\
    \mathrm{IV} &= [(\mathbf{W}_l^T\odot[\mathbf{R}_{y_lq_l}\mathbf{W}^H_l])\mathbf{1}], \\
    \mathrm{V} &= [((\mathbf{R}_{y_lq_l}^*\mathbf{W}_l^T)\odot\mathbf{W}_l^H)\mathbf{1}].
    \end{align}

    Substituting these results in (\ref{eq:deriv_1}) and solving it for
$\mathbf{g}_l$ we obtain
     \begin{align}
     \mathbf{g}_l&=[(\mathbf{W}_l^T\mathbf{W}_l^*) \odot \mathbf{R}_{y_ly_l}+(\mathbf{W}_l^H\mathbf{W}_l) \odot \mathbf{R}_{y_ly_l}^T]^{-1} \nonumber \\
     &\cdot\frac{2}{\alpha}(\mathrm{Re}([(\mathbf{R}_{x_ly_l}^T\odot\mathbf{W}_l^H)\mathbf{1}])
     - \mathrm{Re}([(\mathbf{W}_l^T\odot(\mathbf{R}_{y_lq_l}\mathbf{W}_l^H))\mathbf{1}])).
     \nonumber
     \label{eq:g_l}
     \end{align}

    In the following we outline the computation of the clipping factor
$\alpha$ based on the received signal power. The received signal
power at the $l$-th RRH can be computed by
     \begin{eqnarray}
     P_l &=&\tr(\mathbf{R}_{y_ly_l}+\mathbf{R}_{y_lq_l}+\mathbf{R}_{y_lq_l}^H+\mathbf{R}_{q_lq_l}),
     \end{eqnarray}
\noindent and the received symbol energy by
     \begin{eqnarray}
     E_{rx}=\sqrt{\frac{ \tr(\mathbf{R}_{y_ly_l}+\mathbf{R}_{y_lq_l}+\mathbf{R}_{y_lq_l}^H+\mathbf{R}_{q_lq_l})}{N_R}}.
     \end{eqnarray}

    Thus, the clipping factor $\alpha$ can be obtained from
     \begin{eqnarray}
     \alpha=\gamma\sqrt{\frac{ \tr(\mathbf{R}_{y_ly_l}+\mathbf{R}_{y_lq_l}+\mathbf{R}_{y_lq_l}^H+\mathbf{R}_{q_lq_l})}{N_R}},
     \end{eqnarray}
\noindent where $\gamma$ is a calibration factor. To ensure an
optimized performance, the value of $\gamma$ was set to
$\frac{\sqrt{b}}{2}$, which corresponds to the modulus of the last
quantizer label.
\vspace{-12pt}
\subsection{LRA-MMSE linear receiver in the CU}

We assume that the received signals at each RRH are processed by
independent AGCs before quantization. After the quantization, both the output signals of the ADCs and the optimized AGC coefficients are sent to the CU. Then, the transmitted signal vector by the $l$-th RRH is expressed by
\begin{eqnarray}
\mathbf{r}_l=\mathrm{Q}(\mathrm{diag}(\mathbf{g}_l)\mathbf{y}_l)=\mathrm{diag}(\mathbf{g}_l)\mathbf{y}_l+\mathbf{q}_l.
\label{eq:sent_received_CU}
\end{eqnarray}

    In many prior works related to large-scale MIMO and C-RANs, it has been assumed that the receivers are connected to the CU via error-free FH links. However, those assumptions are unrealistic in practical systems. In this work, we assume that the AGC coefficients are sent to the CU through high-resolution control channels of imperfect FH links. Then, those coefficients arrive at the CU with the additive noise due to transmission errors from the FH links. At the CU the received vector with the AGC coefficients from all RRHs of the cluster can be written as
    \begin{eqnarray}
    \tilde{\mathbf{g}}=\mathbf{g}+\mathbf{n}_{\mathrm{FH}},
    \label{eq:agc_cu}
    \end{eqnarray}
\noindent where $\mathbf{g}=[\mathbf{g}_1^T,...,\mathbf{g}_L^T]^T$ is the vector with the coefficients of all AGCs of the cluster and $\mathbf{n}_{\mathrm{FH}}\in \mathbb{C}^{LN_R \times 1}$ is the vector that models the noise that corrupts the FH transmissions and leads to errors. The elements of $\mathbf{n}_{\mathrm{FH}}$ are considered to be i.i.d ZMCSCG random variables with variance $\sigma_{\tau}^2$.
    From (\ref{eq:sent_received_CU}) and (\ref{eq:agc_cu}) the digital signal vector at the CU can be expressed as
\begin{eqnarray}
\mathbf{r}_{\mathrm{cu}}  &=& \tilde{\mathbf{G}}(\hat{\mathbf{H}}\mathbf{x}+\mathbf{n})+\mathbf{q} = \tilde{\mathbf{G}}\mathbf{y}+\mathbf{q},
\label{eq:received_signal_cluster}
\end{eqnarray}
\noindent where $\tilde{\mathbf{G}}=\diag(\tilde{\mathbf{g}})$, $\mathbf{n} \in \mathbb{C}^{LN_R \times 1}$ contains
AWGN samples, $\mathbf{q} \in \mathbb{C}^{LN_R \times 1}$ is the
quantization noise vector, and  $\hat{\mathbf{H}} \in \mathbb{C}^{LN_R \times
LKN_T}$ contains the channel coefficients between all users and all RRHs of the cluster. The linear receiver $\mathbf{W}_{LRA}$ that minimizes the MSE cost function
\begin{eqnarray}
\varepsilon =\mathrm{E}[||\mathbf{x}-\hat{\mathbf{x}}||^2_2]=\mathrm{E}[||\mathbf{x}-\mathbf{W}_{LRA}\mathbf{r}_{\mathrm{cu}}||^2_2],
\end{eqnarray}

\noindent is given by
\begin{eqnarray}
\mathbf{W}_{LRA}=\mathbf{R}_{xr_{\mathrm{cu}}}\mathbf{R}_{r_{\mathrm{cu}}r_{\mathrm{cu}}}^{-1},
\label{eq:wiener2}
\end{eqnarray}
\noindent where the cross-correlation matrix $\mathbf{R}_{xr_{\mathrm{cu}}} \in \mathbb{C}^{
LKN_T \times LN_R}$, and the autocorrelation matrix $\mathbf{R}_{r_{\mathrm{cu}}r_{\mathrm{cu}}} \in \mathbb{C}^{LN_R \times LN_R}$ are given by
\begin{eqnarray}
\mathbf{R}_{xr_{\mathrm{cu}}}&=&\mathbf{R}_{xy}\tilde{\mathbf{G}}+\mathbf{R}_{xq}, \nonumber \\
\mathbf{R}_{r_{\mathrm{cu}}r_{\mathrm{cu}}}&=&\hspace{-5pt}\tilde{\mathbf{G}}\mathbf{R}_{yy}\tilde{\mathbf{G}}+\tilde{\mathbf{G}}\mathbf{R}_{yq}+\mathbf{R}_{yq}^H\tilde{\mathbf{G}}+\mathbf{R}_{qq}.\nonumber
\end{eqnarray}

    During the detection process, the $i$-th estimated symbol $\hat{x}_i$ from the estimated signal vector $\hat{\mathbf{x}}$, is defined as $\mathrm{DEC}(\mathbf{w}_{LRA}(i,:)\mathrm{Q}(\tilde{\mathbf{G}}\mathbf{y}))$, where $\mathrm{DEC}(\cdot)$ is the slicer function appropriate for the modulation scheme being used in the system. From a given constellation alphabet $\mathbb{X}$, this operation chooses the constellation point with smallest Euclidean distance to the estimated symbol,
\vspace{-4pt}
\begin{eqnarray}
\hat{x}_i=\underset{x \in \mathbb{X}}{\mathrm{argmin}} ||\mathbf{W}_{LRA}(i,:)\mathrm{Q}(\tilde{\mathbf{G}}\mathbf{y})-x||.
\end{eqnarray}

    In Algorithm \ref{alg:linear_filter} we detail the procedure of how
to obtain the AGC of each cell, the LRA-MMSE linear receive filter
in the CU and the linear detection scheme.
\vspace{-5pt}
\begin{algorithm}[H]
\caption{AGC-LRA-MMSE linear receiver}
\begin{algorithmic}[1]
        \State Initialize parameters $\rho_q,\beta, \mathbf{y}, \mathbf{R}_{xx}, \mathbf{R}_{x_lx_l}, \mathbf{R}_{nn}, \mathbf{R}_{n_ln_l};$
        \For{$l = 1$ to $L$}
            \State $\mathbf{R}_{y_ly_l}=\hat{\mathbf{H}}_l\mathbf{R}_{xx}\hat{\mathbf{H}}_l^H+\mathbf{R}_{n_ln_l};$
            \State $\mathbf{R}_{x_ly_l}=\mathbf{R}_{x_lx_l}\hat{\mathbf{H}}_{ll}^H;$
            \State $\mathbf{R}_{y_lq_l}=-\rho_q\mathbf{R}_{y_ly_l};$
            \State $\mathbf{R}_{q_lq_l}=\rho_q\mathbf{R}_{y_ly_l}-(1-\rho_q)\rho_q \nondiag(\mathbf{R}_{y_ly_l});$
            \State $\mathbf{W}_l=\mathbf{R}_{x_ly_l}(\mathbf{R}_{y_ly_l}-\rho_q\nondiag(\mathbf{R}_{y_ly_l}))^{-1};$
            \State $\alpha=\gamma \cdot\sqrt{ \tr(\mathbf{R}_{y_ly_l}+\mathbf{R}_{y_lq_l}+\mathbf{R}_{y_lq_l}^H+\mathbf{R}_{q_lq_l})/(N_R)};$
            \State $\mathbf{g}_l=[(\mathbf{W}_l^T\mathbf{W}_l^*) \odot \mathbf{R}_{y_ly_l}+(\mathbf{W}_l^H\mathbf{W}_l) \odot \mathbf{R}_{y_ly_l}^T]^{-1}\cdot\frac{2}{\alpha}(\mathrm{Re}([(\mathbf{R}_{x_ly_l}^T\odot\mathbf{W}_l^H)\mathbf{1}])
     - \mathrm{Re}([(\mathbf{W}_l^T\odot(\mathbf{R}_{y_lq_l}\mathbf{W}_l^H))\mathbf{1}]));$
        \EndFor
        \State  $\tilde{\mathbf{G}}=\diag([\tilde{\mathbf{g}}_1^T,...,\tilde{\mathbf{g}}_l^T,...,\tilde{\mathbf{g}}_L^T]^T);$
        \State  $\mathbf{R}_{yy}=\hat{\mathbf{H}}\mathbf{R}_{xx}\hat{\mathbf{H}}^H+\mathbf{R}_{nn};$
        \State  $\mathbf{R}_{xy}=\mathbf{R}_{xx}\hat{\mathbf{H}}^H;$
        \State  $\mathbf{R}_{xq}=-\rho_q\mathbf{R}_{xy};$
        \State  $\mathbf{R}_{yq}=\mathbf{R}_{yy}-(1-\rho_q)\rho_q \nondiag(\mathbf{R}_{yy});$
        \State  $\mathbf{R}_{qq}=\rho_q \mathbf{R}_{yy}-(1-\rho_q)\rho_q \nondiag(\mathbf{R}_{yy});$
        \State  $\mathbf{R}_{r_{\mathrm{cu}}r_{\mathrm{cu}}}=\tilde{\mathbf{G}}\mathbf{R}_{yy}\tilde{\mathbf{G}}+\tilde{\mathbf{G}}\mathbf{R}_{yq}+\mathbf{R}_{yq}^H\tilde{\mathbf{G}}+\mathbf{R}_{qq};$
        \State  $\mathbf{R}_{xr_{\mathrm{cu}}}=\mathbf{R}_{xy}\tilde{\mathbf{G}}+\mathbf{R}_{xq};$
        \State  $\mathbf{W}_{LRA}=\mathbf{R}_{xr_{\mathrm{cu}}}\mathbf{R}_{r_{\mathrm{cu}}r_{\mathrm{cu}}}^{-1};$
        \State  $\hat{\mathbf{x}}=\mathrm{DEC}(\mathbf{W}_{LRA}\mathrm{Q}(\tilde{\mathbf{G}}\mathbf{y}));$
\end{algorithmic}
\label{alg:linear_filter}
\end{algorithm}
\vspace{-20pt}
\subsection{LRA-MMSE-SIC receiver in the CU}
\vspace{-5pt}
    SIC detectors can outperform linear detectors and achieve the
sum-capacity in the uplink of multiuser MIMO systems \cite{D_tse}.
At each time, a data stream is decoded and its contribution is
removed from the received signal. SIC detectors improve the signal-to-interference-plus-noise ratio (SINR) of the remaining symbols that will be detected in the following stages and consequently improve the detection accuracy. Unfortunately, SIC techniques suffer from error propagation. To improve the performance of the SIC detector, in this
work data streams are ordered based on channel powers
\cite{sic_1,sic_2}.

    In the BBU pool the received signal at the
$a$-th stage of a SIC detector, $\mathbf{y}^{(a)} \in
\mathbb{C}^{LN_R \times 1}$, is given by
    \begin{eqnarray}
    \mathbf{y}^{(a)}=
    \begin{cases}
    \mathbf{y}^{(1)}, & a=1,\\
    \mathbf{y}^{(1)}-\displaystyle \sum_{j=1}^{a-1}\mathbf{h}^{\mathbf{\Phi}(j)}\hat{x}^{\mathbf{\Phi}(j)}, & 2\leqslant a \leqslant LKN_T,
    \end{cases}
    \end{eqnarray}
\noindent where $\hat{x}^{\mathbf{\Phi}(j)}$ is the symbol estimated
in the $j$-th stage prior to the $a$-th stage and
$\mathbf{h}^{\mathbf{\Phi}(j)} \in \mathbb{C}^{LN_R \times
1}$ is the $\mathbf{\Phi}(j)$-th column of  $\hat{\mathbf{H}}$. In
this notation, $\mathbf{\Phi}$ corresponds to the ordering vector,
whose entries indicate what is the symbol that has to be detected at each stage. After detection, the corresponding column
$\mathbf{h}^{\mathbf{\Phi}(a)}$ from the channel matrix
$\hat{\mathbf{H}}^{(a)} \in \mathbb{C}^{LN_R \times (LKN_T-a+1)}$
is cancelled and another LRA-MMSE receive filter is computed for the
next stage. The quantized received signal vector $\mathbf{r}^{(a)}
\in \mathbb{C}^{LN_R \times 1}$ at the $a$-th stage is given by
\begin{eqnarray}
\mathbf{r}^{(a)}_{\mathrm{cu}} = \mbox{Q}(\tilde{\mathbf{G}}\mathbf{y}^{(a)}) =  \tilde{\mathbf{G}}(\hat{\mathbf{H}}^{(a)}\mathbf{x}^{(a)}+\mathbf{n})+\mathbf{q}^{(a)}.
\end{eqnarray}
The LRA-MMSE linear receive filter of the SIC detector is given by
\vspace{-6pt}
    \begin{eqnarray}
    \mathbf{W}_{LRA}^{(a)}=\mathbf{R}_{xr_{\mathrm{CU}}}^{(a)}(\mathbf{R}_{r_{\mathrm{CU}}r_{\mathrm{CU}}}^{(a)})^{-1},
    \end{eqnarray}
\noindent where the cross-correlation matrix $\mathbf{R}_{xr}^{(a)} \in \mathbb{C}^{(LKN_T-a+1) \times LN_R}$ and the autocorrelation matrix $\mathbf{R}_{rr}^{(a)} \in \mathbb{C}^{LN_R \times LN_R}$ are given by
\begin{eqnarray}
\mathbf{R}_{xr_{\mathrm{CU}}}^{(a)}&=&\mathbf{R}_{xy}^{(a)}\tilde{\mathbf{G}}+\mathbf{R}_{xq}^{(a)}, \label{eq:Rxr_wiener2} \\
\mathbf{R}_{r_{\mathrm{CU}}r_{\mathrm{CU}}}^{(a)}&=&\tilde{\mathbf{G}}\mathbf{R}_{yy}^{(a)}\tilde{\mathbf{G}}+\tilde{\mathbf{G}}\mathbf{R}_{yq}^{(a)}+(\mathbf{R}_{yq}^{(a)})^H\tilde{\mathbf{G}}\nonumber \\
&&+\mathbf{R}_{qq}^{(a)}.\label{eq:Rrr_wiener2}
\end{eqnarray}

    The joint AGC and LRA-MMSE linear
receive filter design with SIC detection scheme is illustrated in
Algorithm \ref{alg:sic_filter}.
\vspace{-12pt}

\begin{algorithm}[H]
\caption{AGC-LRA-MMSE with SIC receiver}
\begin{algorithmic}[1]
        \State Initialize parameters $\rho_q,\beta, \mathbf{y}, \mathbf{R}_{xx}, \mathbf{R}_{x_lx_l}, \mathbf{R}_{nn}, \mathbf{R}_{n_ln_l};$
        \State Ordering $\mathbf{\Phi}=[\Phi_1,\Phi_2,...,\Phi_{LKN_T}];$
        \For{$l = 1$ to $L$}
            \State $\mathbf{R}_{y_ly_l}=\hat{\mathbf{H}}_l\mathbf{R}_{xx}\hat{\mathbf{H}}_l^H+\mathbf{R}_{n_ln_l};$
            \State $\mathbf{R}_{x_ly_l}=\mathbf{R}_{x_lx_l}\hat{\mathbf{H}}_{ll}^H;$
            \State $\mathbf{R}_{y_lq_l}=-\rho_q\mathbf{R}_{y_ly_l};$
            \State $\mathbf{R}_{q_lq_l}=\rho_q\mathbf{R}_{y_ly_l}-(1-\rho_q)\rho_q \nondiag(\mathbf{R}_{y_ly_l});$
            \State $\mathbf{W}_l=\mathbf{R}_{x_ly_l}(\mathbf{R}_{y_ly_l}-\rho_q\nondiag(\mathbf{R}_{y_ly_l}))^{-1};$
            \State $\alpha=\gamma \cdot\sqrt{ \tr(\mathbf{R}_{y_ly_l}+\mathbf{R}_{y_lq_l}+\mathbf{R}_{y_lq_l}^H+\mathbf{R}_{q_lq_l})/(N_R)};$
            \State $\mathbf{g}_l=[(\mathbf{W}_l^T\mathbf{W}_l^*) \odot \mathbf{R}_{y_ly_l}+(\mathbf{W}_l^H\mathbf{W}_l) \odot \mathbf{R}_{y_ly_l}^T]^{-1}\cdot\frac{2}{\alpha}(Re([(\mathbf{R}_{x_ly_l}^T\odot\mathbf{W}_l^H)\mathbf{1}])
     - Re([(\mathbf{W}_l^T\odot(\mathbf{R}_{y_lq_l}\mathbf{W}_l^H))\mathbf{1}]));$
        \EndFor
        \State  $\tilde{\mathbf{G}}=\diag([\tilde{\mathbf{g}}_1^T,...,\tilde{\mathbf{g}}_l^T,...,\tilde{\mathbf{g}}_L^T]^T);$
        \For{$a = 1$ to $LKN_T$}
        \State  $\mathbf{R}_{yy}=\hat{\mathbf{H}}\mathbf{R}_{xx}\hat{\mathbf{H}}^H+\mathbf{R}_{nn};$
        \State  $\mathbf{R}_{xy}=\mathbf{R}_{xx}\hat{\mathbf{H}}^H;$
        \State  $\mathbf{R}_{xq}=-\rho_q\mathbf{R}_{xy};$
        \State  $\mathbf{R}_{yq}=\mathbf{R}_{yy}-(1-\rho_q)\rho_q \nondiag(\mathbf{R}_{yy});$
        \State  $\mathbf{R}_{qq}=\rho_q \mathbf{R}_{yy}-(1-\rho_q)\rho_q \nondiag(\mathbf{R}_{yy});$
        \State  $\mathbf{R}_{r_{\mathrm{cu}}r_{\mathrm{cu}}}=\tilde{\mathbf{G}}\mathbf{R}_{yy}\tilde{\mathbf{G}}+\tilde{\mathbf{G}}\mathbf{R}_{yq}+\mathbf{R}_{yq}^H\tilde{\mathbf{G}}+\mathbf{R}_{qq};$
        \State  $\mathbf{R}_{xr_{\mathrm{cu}}}=\mathbf{R}_{xy}\tilde{\mathbf{G}}+\mathbf{R}_{xq};$
        \State  $\mathbf{W}_{LRA}=\mathbf{R}_{xr_{\mathrm{cu}}}\mathbf{R}_{r_{\mathrm{cu}}r_{\mathrm{cu}}}^{-1};$
        \State  $\hat{x}=\mathrm{DEC}(\mathbf{W}_{LRA}(\mathbf{\Phi}(a),:)\mbox{Q}(\tilde{\mathbf{G}}\mathbf{y}));$
        \State  $\mathbf{y}=\mathbf{y}-\mathbf{H}(:,\mathbf{\Phi}(a))\hat{x};$
        \State  $\mathbf{H}(:,\mathbf{\Phi}(a))=\zeros(LN_R,1);$
        \EndFor
\end{algorithmic}
\label{alg:sic_filter}
\end{algorithm}
\subsection{Computational Complexity}

    The computational complexity of the proposed AGC-LRA-MMSE linear and
SIC receivers can be exactly computed as a function of the number of
receive and transmit antennas, the
number of users per cell and the number of cells as
depicted in Table \ref{tabela:complexiade}. To assess the
computational complexity of the AGC-LRA-MMSE receivers we have
computed the number of arithmetic operations such as complex
additions and multiplications.

    To initialize the AGC-LRA-MMSE algorithm, a linear receive filter
$\mathbf{W}_l$ is computed for each cell by (\ref{eq:l_rec_cell}).
The largest contribution in terms of computational complexity in the
computation of  $\mathbf{W}_l$ is due to the inversion of
$\mathbf{R}_{r_lr_l} \in \mathbb{C}^{N_R \times N_R}$. In this work
we consider that the inversion of an $N \times N$ matrix by Gaussian
elimination costs $\Or(N^3)$ operations. Therefore, the
computational cost to obtain each $\mathbf{W}_l$ matrix is
$\Or(N_R^3)$. After that, an AGC matrix $\diag(\mathbf{g}_l)$ with a
computational cost of $\Or(N_R^3)$ is computed for each RRH. Then,
an LRA linear receive filter $\mathbf{W}_{LRA}$ is computed by
(\ref{eq:wiener2}), which requires the inversion of the matrix
$\mathbf{R}_{rr} \in \mathbb{C}^{LN_R \times LN_R}$. Thus, the
computational complexity to obtain $\mathbf{W}_{LRA}$ is
$\Or(L^3N_R^3)$. Summarizing these results, the proposed
AGC-LRA-MMSE linear receive filter has a total cost of
$\Or\left(N_R^3(L^3+2L)\right)$. When we employ SIC detection with
the AGC-LRA-MMSE receiver filter an $W_{LRA}^{(a)}$ matrix is
computed to detect a symbol at each stage of the interference
cancellation. Thus, as we consider the transmission of $LKN_T$
streams, the expression of the LRA-MMSE receive filter
$W_{LRA}^{(a)}$ is computed $LKN_T$ times to detect all data
streams. Therefore, the computational complexity of the proposed
AGC-LRA-MMSE-SIC algorithm is $\Or\left(L^4N_R^3KN_T\right)$. We
remark that these costs can be reduced by the efficient use signal
processing algorithms, which can be investigated in a future work.

\begin{table}[H]
\centering
\caption{Computational complexity of algorithms.}
\label{tabela:complexiade}
\begin{tabular}{c l l}
\hline
Task & Additions & Multiplications \\
\hline
\vspace{1pt}
FR Standard MMSE & $\Or(L^3N_R^3)$ &$\Or(L^3N_R^3)$  \vspace{2pt} \\
AGC-LRA-MMSE& $\Or\left(N_R^3(L^3+2L)\right)$ &$\Or\left(N_R^3(L^3+2L)\right)$  \vspace{2pt}\\
AGC-LRA-MMSE-SIC& $\Or\left(L^4N_R^3KN_T\right)$ & $\Or\left(L^4N_R^3KN_T\right)$  \vspace{2pt}\\
\hline
\end{tabular}
\end{table}

\section{Sum rate analysis}
In this section, assuming Gaussian signaling, we derive expressions for the achievable sum rates of the proposed joint AGC and LRA-MMSE receive filter design in large-scale MIMO with C-RAN systems for linear and SIC schemes.
\subsection{Sum Rate of Linear Receivers}

The ergodic sum rate $\mathcal{R}_{\Sum}$ of the system with the
AGC-LRA-MMSE linear receive filter is given by the sum of the
achievable rates of each user in the cluster, averaged over the channel
realizations as described by
\begin{align}
\mathcal{R}_{\Sum}&=\sum_{l=1}^L\sum_{k=1}^K \mathrm{E}_\mathbf{H}[\mathcal{R}_{l}^{(k)}].
\end{align}

    The achievable rate $\mathcal{R}_{l}^{(k)}$ of the $k$-th user in the
$l$-th cell is calculated as
\begin{eqnarray}
\mathcal{R}_{l}^{(k)}=\log_2 \det(\mathbf{I}_{N_T}+\Lambda_l^{(k)}),
\label{eq:rate_userk}
\end{eqnarray}
\noindent where $\Lambda_l^{(k)}$ denotes a matrix associated with
the post processing signal-to-interference-plus-noise ratio (SINR)
of the $k$-th user in the $l$-th cell given by
\begin{eqnarray}
\Lambda_l^{(k)}=\mathbf{\Upsilon}_l^{(k)}(\mathbf{\Gamma}_l^{(k)})^{-1},
\label{eq:94584}
\end{eqnarray}
\noindent where $\mathbf{\Upsilon}_l^{(k)}$ represents the covariance matrix of the desired signal and $\mathbf{\Gamma}_l^{(k)}$ represents the covariance matrix of the noise plus interference \cite{adc_uniform_chosen_1, ergodic,cover}. At the BBU pool, the received signals of the cluster can be computed by (\ref{eq:received_signal_cluster}). Assuming that the BBU pool employs the LRA-MMSE receiver to detect the symbols transmitted  by the users, we can compute the estimated symbol of the $k$-th user at the $l$-th cell by
\begin{align}
\hat{\mathbf{x}}_l^{(k)}&=\mathbf{W}_{LRA,l}^{(k)}\tilde{\mathbf{G}}\hat{\mathbf{H}}_{l}^{(k)}\mathbf{x}_l^{(k)}+\sum_{\stackrel{u=1}{u\neq k}}^{K}\mathbf{W}_{LRA,l}^{(k)}\tilde{\mathbf{G}}\hat{\mathbf{H}}_{l}^{(u)}\mathbf{x}_l^{(u)}\nonumber \\
&+\sum_{\stackrel{j=1}{j\neq l}}^{L}\sum_{u=1}^{K}\mathbf{W}_{LRA,l}^{(k)}\tilde{\mathbf{G}}\hat{\mathbf{H}}_{j}^{(u)}\mathbf{x}_j^{(u)}+\mathbf{W}_{LRA,l}^{(k)}\tilde{\mathbf{G}}\mathbf{n} \nonumber \\
&+\mathbf{W}_{LRA,l}^{(k)}\mathbf{q}.
\label{eq:est_symb_user}
\end{align}

    In (\ref{eq:est_symb_user}), the first term corresponds to the
estimation of the desired symbol and the other terms are
interferences. Thus, the covariance matrix of the desired signal is
given by
\begin{eqnarray}
\mathbf{\Upsilon}_l^{(k)}=\sigma_x^2(\mathbf{W}_{LRA,l}^{(k)}\tilde{\mathbf{G}}\hat{\mathbf{H}}_{l}^{(k)})(\mathbf{W}_{LRA,l}^{(k)}\tilde{\mathbf{G}}\hat{\mathbf{H}}_{l}^{(k)})^H.
\label{eq:1678}
\end{eqnarray}
    The other terms of (\ref{eq:est_symb_user}) are the interferences
present in the system such as the intracell interference, the
intercell interference, the AWGN and the quantization distortion.
Therefore, the covariance matrix of the noise plus interference part can be obtained by
\begin{align}
\mathbf{\Gamma}_l^{(k)}=&\sigma_x^2\sum_{\stackrel{u=1}{u\neq k}}^{K}(\mathbf{W}_{LRA,l}^{(k)}\tilde{\mathbf{G}}\hat{\mathbf{H}}_{l}^{(u)})(\mathbf{W}_{LRA,l}^{(k)}\tilde{\mathbf{G}}\hat{\mathbf{H}}_{l}^{(u)})^H\nonumber \\
&-\rho_q \sigma_x^2\sum_{\stackrel{u=1}{u\neq k}}^{K}(\mathbf{W}_{LRA,l}^{(k)}\hat{\mathbf{H}}_{l}^{(u)})(\mathbf{W}_{LRA,l}^{(k)}\tilde{\mathbf{G}}\hat{\mathbf{H}}_{l}^{(u)})^H \nonumber \\
&-\rho_q \sigma_x^2\sum_{\stackrel{u=1}{u\neq k}}^{K}(\mathbf{W}_{LRA,l}^{(k)}\tilde{\mathbf{G}}\hat{\mathbf{H}}_{l}^{(u)})(\mathbf{W}_{LRA,l}^{(k)}\hat{\mathbf{H}}_{l}^{(u)})^H\nonumber \\
&+\sigma_x^2 \sum_{\stackrel{j=1}{i\neq l}}^{L}\sum_{u=1}^{K}(\mathbf{W}_{LRA,l}^{(k)}\tilde{\mathbf{G}}\hat{\mathbf{H}}_{j}^{(u)})(\mathbf{W}_{LRA,l}^{(k)}\tilde{\mathbf{G}}\hat{\mathbf{H}}_{j}^{(u)})^H \nonumber \\
&-\rho_q \sigma_x^2\sum_{\stackrel{i=1}{i\neq l}}^{L}\sum_{u=1}^{K}(\mathbf{W}_{LRA,l}^{(k)}\hat{\mathbf{H}}_{j}^{(u)})(\mathbf{W}_{LRA,l}^{(k)}\tilde{\mathbf{G}}\hat{\mathbf{H}}_{j}^{(u)})^H\nonumber \\
&-\rho_q \sigma_x^2\sum_{\stackrel{i=1}{i\neq l}}^{L}\sum_{u=1}^{K}(\mathbf{W}_{LRA,l}^{(k)}\tilde{\mathbf{G}}\hat{\mathbf{H}}_{j}^{(u)})(\mathbf{W}_{LRA,l}^{(k)}\hat{\mathbf{H}}_{j}^{(u)})^H \nonumber \\
&+\sigma_n^2(\mathbf{W}_{LRA,l}^{(k)}\tilde{\mathbf{G}})(\mathbf{W}_{LRA,l}^{(k)}\tilde{\mathbf{G}})^H
\nonumber \\
&-\rho_q\sigma_n^2(\mathbf{W}_{LRA,l}^{(k)})(\mathbf{W}_{LRA,l}^{(k)}\tilde{\mathbf{G}})^H
\nonumber \\
&-\rho_q\sigma_n^2(\mathbf{W}_{LRA,l}^{(k)}\tilde{\mathbf{G}})(\mathbf{W}_{LRA,l}^{(k)})^H +\mathbf{W}_{LRA,l}^{(k)}\rho_q \cdot\nonumber \\
&(\mathbf{R}_{yy}-(1-\rho_q) \nondiag(\mathbf{R}_{yy}))\mathbf{W}_{LRA,l}^{(k)H}.
\label{eq:15605}
\end{align}

    Substituting (\ref{eq:1678}) and (\ref{eq:15605}) in
(\ref{eq:94584}) we get the expression of the matrix associated with
the post processing SINR of the $k$-th user in the $l$-th cell. Then,
we can substitute (\ref{eq:94584}) in (\ref{eq:rate_userk}) to get
the achievable rate $\mathcal{R}_{l}^{(k)}$. Using (\ref{eq:rate_userk}) the ergodic
sum rate $\mathcal{R}_{\Sum}$ of the system, averaged over
channel realizations, is described by
\begin{align}
\mathcal{R}_{\Sum}=\sum_{l=1}^L\sum_{k=1}^K \mathrm{E}_\mathbf{H}[\log_2 \det(\mathbf{I}_{N_T}+\mathbf{\Upsilon}_l^{(k)}(\mathbf{\Gamma}_l^{(k)})^{-1})].
\end{align}

\subsection{Sum Rate of SIC receivers}

The uplink sum rate of the SIC receivers based on the proposed
AGC-LRA-MMSE design in a system with $LKN_T$ interfering layers can be computed by the sum of the achievable rate of the $a$-th stream after
the linear receiver with the AGC-LRA-MMSE design, and the achievable
rate of the reduced size $(LKN_T-a) \times LN_R$ MIMO system after
removal of the $a$-th stream, given by
    \begin{align}
\mathcal{R}_{\Sum}=\sum_{a=1}^{MKN_T} \mathrm{E}_\mathbf{H}\left[\log_2\left(1+\frac{\Upsilon^{\mathbf{\Phi}(a)}}{\Gamma^{\mathbf{\Phi}(a)}}\right)\right],
\label{eq:sum_sic}
\end{align}
\noindent where $\Upsilon^{\mathbf{\Phi}(a)}$ is the desired signal
power and $\Gamma^{\mathbf{\Phi}(a)}$ is the interference-plus-noise
power. The expectation is taken over the channel coefficients. In the $a$-th stage, the estimated symbol is given by
    \begin{align}
\hat{x}_l^{\mathbf{\Phi}(a)}=&\mathbf{w}_{LRA,l}^{\mathbf{\Phi}(a)}\tilde{\mathbf{G}}\hat{\mathbf{h}}_{l}^{\mathbf{\Phi}(a)}x_l^{\mathbf{\Phi}(a)}+\sum_{\stackrel{u=1}{u\neq \mathbf{\Phi}(a)}}^{KN_T}\mathbf{w}_{LRA,l}^{\mathbf{\Phi}(a)}\tilde{\mathbf{G}}\hat{\mathbf{h}}_{l}^{(u)}x_l^{(u)}\nonumber \\
&+\sum_{\stackrel{j=1}{j\neq l}}^{L}\sum_{u=1}^{KN_T}\mathbf{w}_{LRA,l}^{\mathbf{\Phi}(a)}\tilde{\mathbf{G}}\hat{\mathbf{h}}_{j}^{(u)}x_j^{(u)}+\mathbf{w}_{LRA,l}^{\mathbf{\Phi}(a)}\tilde{\mathbf{G}}\mathbf{n} \nonumber \\
&+\mathbf{w}_{LRA,l}^{\mathbf{\Phi}(a)}\mathbf{q}, \hspace{15pt} \{x_l^{\mathbf{\Phi}(a)},x_l^{(u)},x_j^{(u)}\} \not\subset \mathbf{\Omega},
\end{align}
\noindent where $\mathbf{\Omega}$ is a set of symbols estimated at
prior stages. The coefficients of the receive filter
$\mathbf{w}_{LRA,l}^{\mathbf{\Phi}(a)}$ are obtained from the
$\mathbf{\Phi}(a)$-th row of the filter matrix
$\mathbf{W}_{LRA}^{(a)}$. Given a channel realization
$\hat{\mathbf{H}}$, the desired signal power is computed by
\begin{eqnarray}
\Upsilon^{\mathbf{\Phi}(a)}=\sigma_x^2(\mathbf{w}_{LRA,l}^{\mathbf{\Phi}(a)}\tilde{\mathbf{G}}\hat{\mathbf{h}}_{l}^{\mathbf{\Phi}(a)})(\mathbf{w}_{LRA,l}^{\mathbf{\Phi}(a)}\tilde{\mathbf{G}}\hat{\mathbf{h}}_{l}^{\mathbf{\Phi}(a)})^H,
\label{eq:Upsilon_sic}
\end{eqnarray}
\noindent where $\hat{\mathbf{h}}^{\mathbf{\Phi}(j)}_l$ is the
$\mathbf{\Phi}(j)$-th column of  $\hat{\mathbf{H}}^{(a)}$. Then,
$\hat{\mathbf{h}}_{l}^{\mathbf{\Phi}(a)}$ becomes null and the
interference-plus-noise
power is given by
\begin{align}
\Gamma^{\mathbf{\Phi}(a)}=&\sigma_x^2(\mathbf{w}_{LRA,l}^{\mathbf{\Phi}(a)}\tilde{\mathbf{G}}\hat{\mathbf{H}}^{(a)})(\mathbf{w}_{LRA,l}^{\mathbf{\Phi}(a)}\tilde{\mathbf{G}}\hat{\mathbf{H}}^{(a)})^H\nonumber \\
&-\rho_q \sigma_x^2[(\mathbf{w}_{LRA,l}^{\mathbf{\Phi}(a)}\hat{\mathbf{H}}_{l}^{(a)})(\mathbf{w}_{LRA,l}^{\mathbf{\Phi}(a)}\tilde{\mathbf{G}}\hat{\mathbf{H}}_{l}^{(a)})^H \nonumber \\
&+(\mathbf{w}_{LRA,l}^{\mathbf{\Phi}(a)}\tilde{\mathbf{G}}\hat{\mathbf{H}}_{l}^{(a)})(\mathbf{w}_{LRA,l}^{\mathbf{\Phi}(a)}\hat{\mathbf{H}}_{l}^{(a)})^H
\nonumber \\
&+\sum_{\stackrel{j=1}{j\neq l}}^{L}(\mathbf{w}_{LRA,l}^{\mathbf{\Phi}(a)}\hat{\mathbf{H}}_{j}^{(a)})(\mathbf{w}_{LRA,l}^{\mathbf{\Phi}(a)}\tilde{\mathbf{G}}\hat{\mathbf{H}}_{j}^{(a)})^H  \nonumber \\
&+\sum_{\stackrel{j=1}{j\neq l}}^{L}(\mathbf{w}_{LRA,l}^{\mathbf{\Phi}(a)}\tilde{\mathbf{G}}\hat{\mathbf{H}}_{j}^{(a)})(\mathbf{w}_{LRA,l}^{\mathbf{\Phi}(a)}\hat{\mathbf{H}}_{j}^{(a)})^H]\nonumber \\
&+\sigma_n^2(\mathbf{w}_{LRA,l}^{\mathbf{\Phi}(a)}\tilde{\mathbf{G}})(\mathbf{w}_{LRA,l}^{\mathbf{\Phi}(a)}\tilde{\mathbf{G}})^H\nonumber \\
&-\rho_q\sigma_n^2[(\mathbf{w}_{LRA,l}^{\mathbf{\Phi}(a)})(\mathbf{w}_{LRA,l}^{\mathbf{\Phi}(a)}\tilde{\mathbf{G}})^H\nonumber \\
&+(\mathbf{w}_{LRA,l}^{\mathbf{\Phi}(a)}\tilde{\mathbf{G}})(\mathbf{w}_{LRA,l}^{\mathbf{\Phi}(a)})^H]+\mathbf{w}_{LRA}^{\mathbf{\Phi}(a)}\rho_q\cdot\nonumber \\
&(\mathbf{R}_{yy}-(1-\rho_q) \nondiag(\mathbf{R}_{yy}))\mathbf{w}_{LRA}^{\mathbf{\Phi}(a)H},
\label{eq:gamma_sic}
\end{align}
\noindent where $\hat{\mathbf{H}}_{l}^{(a)}$ is the channel matrix
between the users in the $l$-th cell and all receive antennas of the cluster.
Substituting (\ref{eq:Upsilon_sic}) and (\ref{eq:gamma_sic}) in
(\ref{eq:sum_sic}) we get the achievable sum rate
$\mathcal{R}_{\Sum}$ of the system when SIC receivers with the
proposed AGC-LRA-MMSE design are employed.
Since the channels are assummed to be wide-sense stationary and drawn from ergodic processes, then $\mathrm{E}_\mathbf{H}$ may be replaced by a simple average over $LH$ in \eqref{eq:rate_userk} and an average over $MKN_T$ in \eqref{eq:sum_sic}.
As one may realize from \eqref{eq:rate_userk} and~\eqref{eq:sum_sic},
the higher the SINR of the system, the higher is the sum rate for both linear and SIC designs.

\section{Simulation Results}

In this section, we discuss the BER performance and the achievable sum rate associated with the proposed algorithms and compare them with the existing techniques in a large-scale MU-MIMO system where the received signals are detected in C-RANs. For our simulations, we consider the uplink channel of a large-scale MU-MIMO system comprised by one cluster with 4 cells. Each cell contains one centralized RRH equipped with $N_R=64$ receive antennas and $K=8$ users, equipped with $N_T=2$ transmit antennas each. The users are distributed randomly and uniformly over the covered area. Moreover, it is also considered that the RRHs share the same frequency band and the system is perfectly synchronized. Synchronization problems can be considered in an even more real scenario. However, a study of the impact of synchronization is not the goal of this work and thus it is left to a future work. The following results show the performance achieved by using both linear and SIC detection schemes.

The channel model used in the simulations includes fast fading, geometric attenuation, and log-normal shadow fading. The small-scale fading is modeled by a Rayleigh channel whose coefficients are i.i.d complex Gaussian random variables with zero-mean and unit variance. The large-scale fading coefficients are obtained by (\ref{eq:large_scale}), where the path-loss exponent is $\gamma=3.7$, and the shadow-fading standard deviation is $\sigma_{\mathrm{shadow}}=8.0$ dB. We consider a cell radius of one kilometer and the users are distributed in a covered area between a cell-hole radius of 10 meters and the cell edge. For each channel realization, each transmit antenna of each user transmits data packets with $100$ symbols using either quadrature phase-shift keying (QPSK) or 16-ary quadrature amplitude modulation (16-QAM). \textcolor{red}{The results are generated using MATLAB and taking the average of $10^3$ channels for both the sum rate and the BER plots. In addition, for each channel realization we have considered $10^5$ symbol vectors and checked if the number of errors was at least $100$ such that the computation of BER curves is sufficiently accurate.}. In each RRH the received signals are treated by independent AGCs and then quantized by uniform quantizers with $b$-bits of resolution before being sent to the CU. In the presented results, the receivers that employ SIC detection scheme are ordered by the channel norms.

\begin{figure}[H]
\centering
\subfloat[Data transfer.]
{
    \includegraphics[scale=0.18]{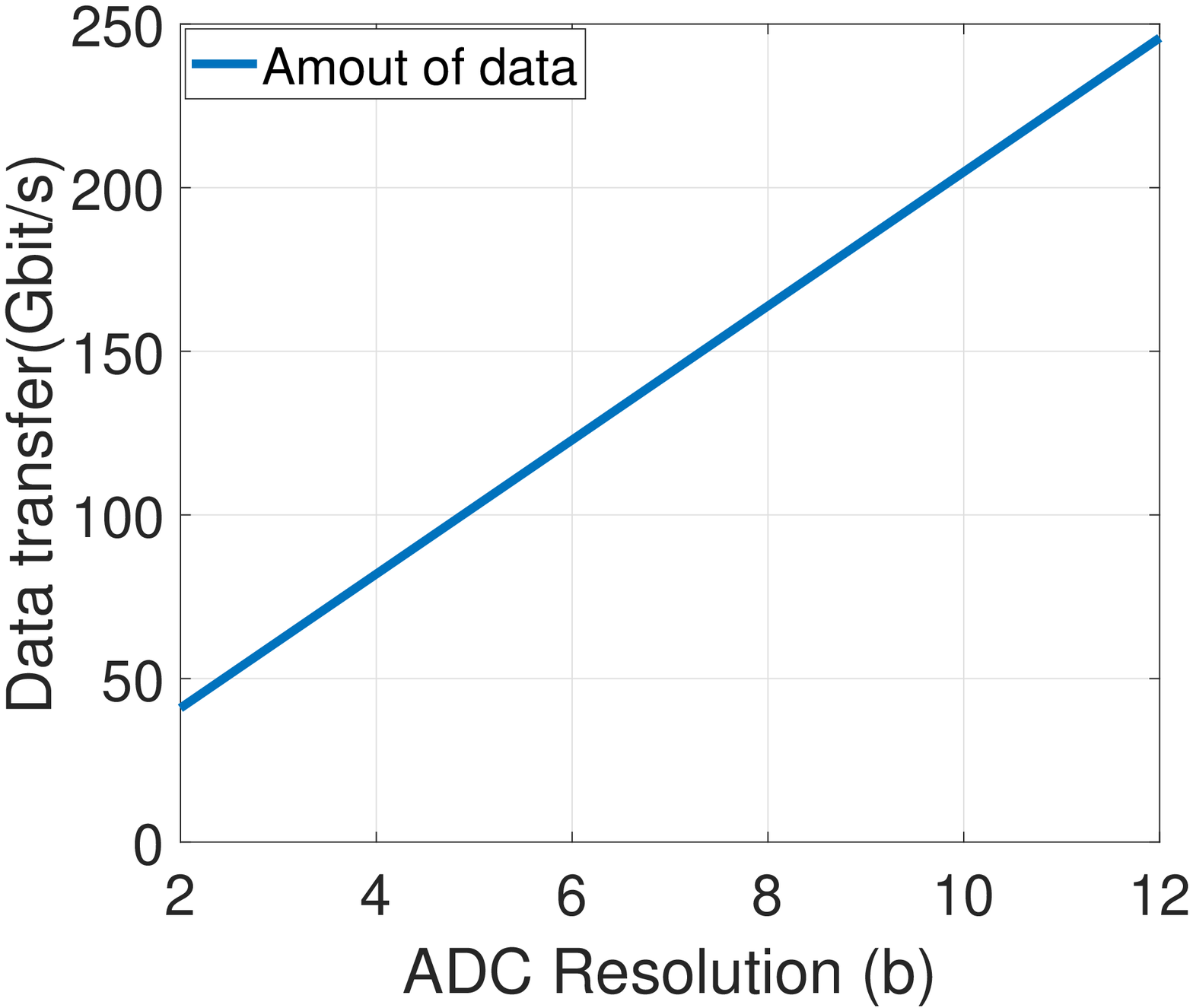}
    \label{fig:amount_of_data}
}
\hfil
\subfloat[Power consumption.]
{
    \includegraphics[scale=0.18]{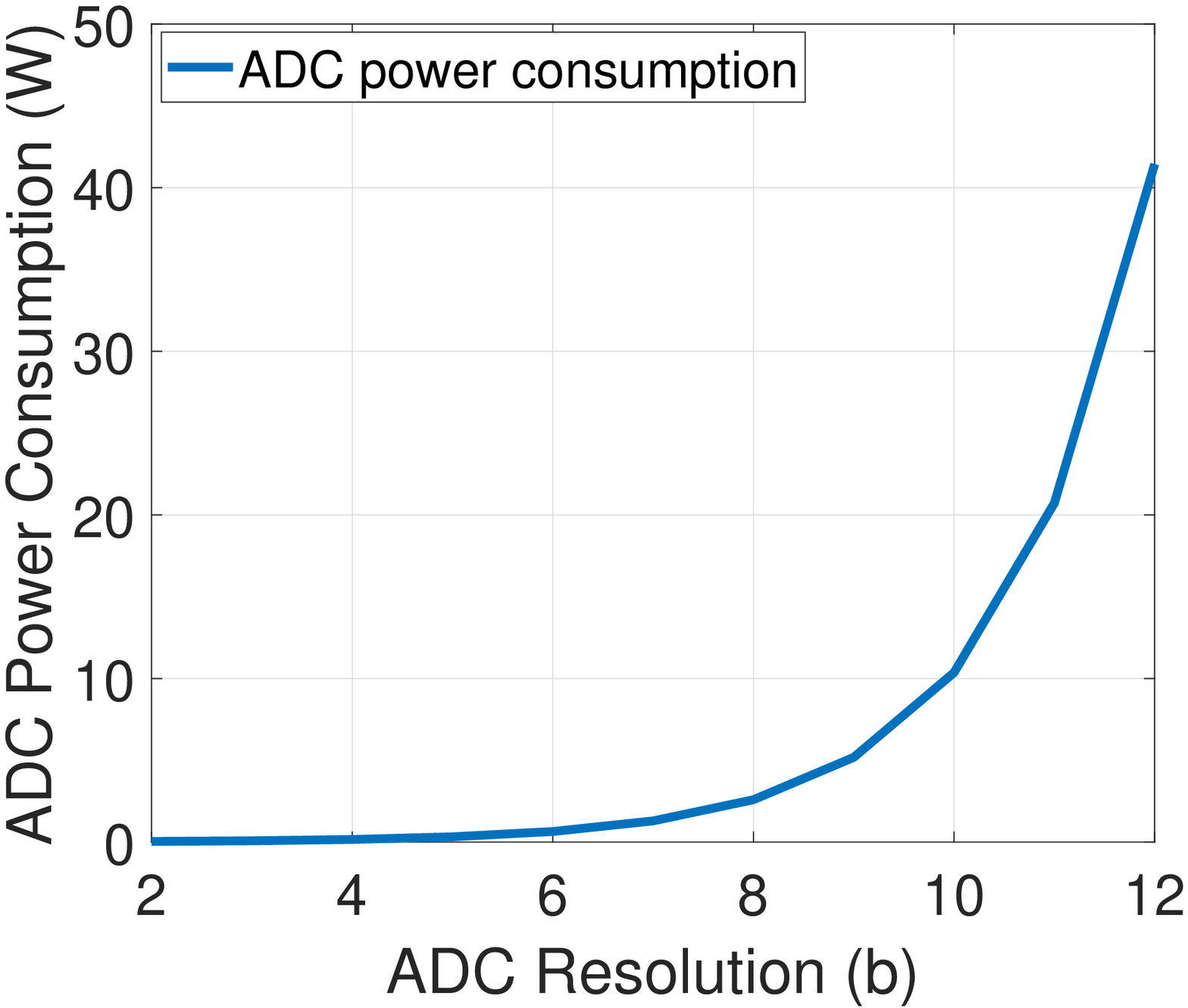}
    \label{fig:power_consumption}
}
\caption{Amount of required data transfer and total ADC power consumption.}
\label{fig:234}
\end{figure}

To highlight how important it is the reduction of the resolution of the ADCs and the achieved gains of the proposed design we illustrate the amount of data that have to be transported through the FH links and the power consumption by the quantization step of the considered scenario. In this paper, we consider that the system bandwidth is equal to $B=20$ MHz. To avoid aliasing we adopt a sampling rate of $W=2B$ in order to satisfy the Nyquist theorem \cite{quant_distort}. The relation between the required data transfer through the FH links and the ADC resolution can be computed as $T(b)=2MN_RWb$, which is illustrated in Fig. \ref{fig:amount_of_data}. The power consumption of each ADC can be calculated as $P_{\mbox{ADC}}(b)=cW2^b$, where $c$ is the power consumption per conversion step (conv-step). This power consumption model of the ADC encompasses various architectures and implementations of ADCs which are described in \cite{59,60,61}. Considering an energy consumption per conversion step $c=494$ fJ \cite{59,61}, the total power consumption by the ADCs is given by $P(b)=2MN_RcW2^b$, which is illustrated, with the proposed example parameters, in Fig. \ref{fig:power_consumption}. The curves of Fig. \ref{fig:234} shows that the deployment of low-resolution techniques can substantially reduce the power consumption from existing solutions that employ high-resolution ADCs (namely 8-12 bits). In particular, the saving in power consumption due to the ADCs in greater than $90\%$. Moreover, the saving in data transfer due the reduction in the number of bits transmitted over the FH links is also significant.

\begin{figure}[H]
\centering
\includegraphics[scale=0.35]{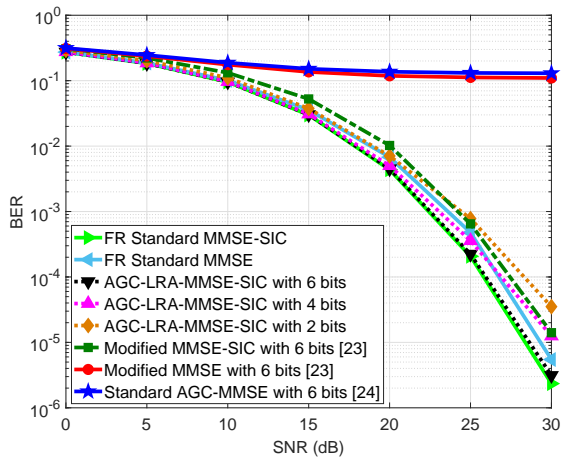}
\caption{Uncoded BER performance comparison with QPSK modulation considering perfect CSIR.}
\label{fig:ber_QPSK}
\end{figure}

In Fig. \ref{fig:ber_QPSK} we investigate the advantages of the proposed AGC-LRA-MMSE receiver design with SIC detection scheme (AGC-LRA-MMSE-SIC) in terms of BER performance when users transmit QPSK symbols. Here we consider that the CSIR is perfectly known and there are not errors in the transmission of the AGC coefficients over the FH links. To investigate the performance gain we consider the Modified MMSE receiver presented in \cite{roth} and the standard AGC from \cite{b2} with the
standard MMSE receiver. For a fair comparison we also employ
\cite{roth} with the SIC detection scheme. The results reveal that, for signals quantized with $6$ bits, the proposed AGC-LRA-MMSE-SIC
approach achieves a very close performance to the performance
achieved by the Full-Resolution (FR) Standard MMSE-SIC receiver in a system with unquantized signals. Moreover, the proposed AGC-LRA-MMSE-SIC detection scheme has a significantly better performance than existing techniques.

\begin{figure}[H]
\centering
\includegraphics[scale=0.475]{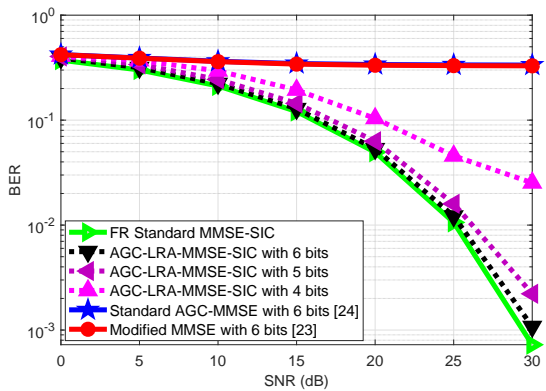}
\caption{Uncoded BER performance comparison with 16-QAM modulation.}
\label{fig:ber_16QAM}
\end{figure}

By increasing the modulation order we consider in Fig. \ref{fig:ber_16QAM} users transmitting 16-QAM modulation symbols. A comparison of Fig. \ref{fig:ber_16QAM} and Fig. \ref{fig:ber_QPSK} shows a significant performance loss due to the higher modulation order. This is expected because the constellation points of the 16-QAM modulation scheme are closer to each other than the constellation points of the QPSK modulation scheme. Thus, the detection of symbols of a higher modulation order is more sensitive to interference. Despite that, Fig. \ref{fig:ber_16QAM} shows a very small gap between the BER achieved by the FR Standard MMSE-SIC scheme and by the proposed AGC-LRA-MMSE-SIC scheme when signals are quantized with $5$ or $6$ bits. Furthermore, we can notice the poor performance achieved by existing techniques when a higher modulation order is considered. The analysis of this result reveals that the proposed AGC-LRA-MMSE-SIC scheme can improve the BER performance even when users transmit symbols of a  higher modulation order.

\vspace{-3em}

\begin{figure}[H]
\centering
\includegraphics[scale=0.35]{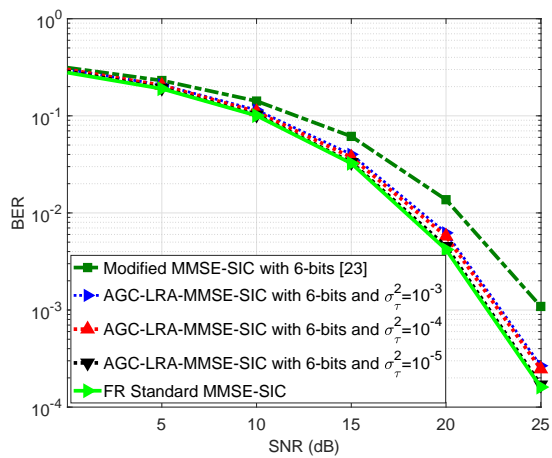}
\caption{AGC-LRA-MMSE-SIC receiver uncoded BER performance comparison with QPSK modulation considering perfect CSIR and $\sigma_{\tau}^2\in\{10^{-3},10^{-4},10^{-5}\}$.}
\label{fig:ber_QPSK_erros_AGC}
\end{figure}

Next we investigate the influence of the imperfect FH links by the possible transmission errors when the optimal AGC coefficients arrive at the CU. In Fig. \ref{fig:ber_QPSK_erros_AGC}  we consider the same setting as in Fig. \ref{fig:ber_QPSK}, but now we take into account the FH transmission errors model as presented by (\ref{eq:agc_cu}). This result shows that even in the presence of errors in the AGC coefficients, the proposed scheme achieves a BER performance close to the FR Standard MMSE-SIC receiver and achieves a better performance than existing techniques.

\begin{figure}[H]
\centering
\includegraphics[scale=0.35]{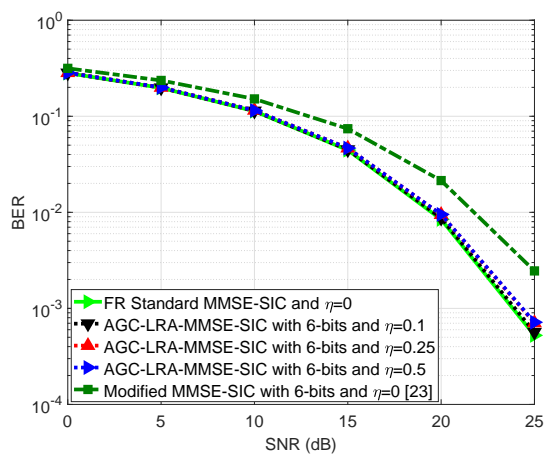}
\caption{AGC-LRA-MMSE-SIC receiver uncoded BER performance comparison with QPSK modulation considering imperfect CSIR with $\eta\in\{0.1, 0.25, 0.5\}$.}
\label{fig:ber_QPSK_erros_channel}
\end{figure}

In order to investigate the BER performance of the proposed AGC-LRA-MMSE-SIC receiver in a system without perfect CSIR we consider the imperfect CSI model described by (\ref{eq:erros_estimacao_canal}). Fig. \ref{fig:ber_QPSK_erros_channel} illustrates the BER performance achieved by the receiver algorithms in a scenario with the CSI related parameter $\eta \in \{0.1, 0.25, 0.5\}$. This result shows the close performance achieved by the AGC-LRA-MMSE-SIC receiver in a scenario whose signals are quantized with 6 bits, and with imperfect CSI, to the performance achieved by the FR-MMSE-SIC receiver with unquantized signals and with perfect CSI. This result confirms that there are no convergence problems in the proposed joint optimization of the AGC and the LRA-MMSE receiver design when the channel is imperfectly known. Moreover, the proposed AGC-LRA-MMSE-SIC receiver still has a better performance than existing techniques even with channel estimation errors.

\begin{figure}[H]
\centering
\includegraphics[scale=0.35]{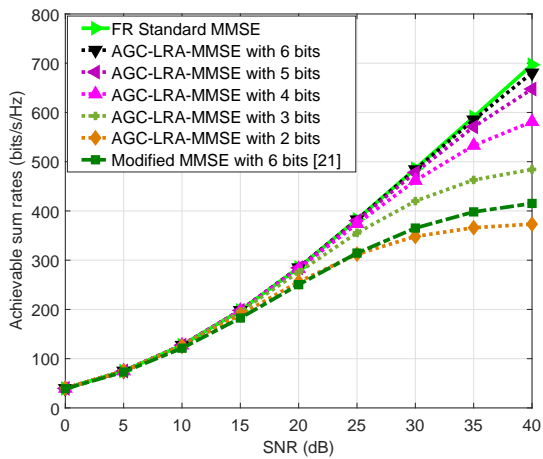}
\caption{Achievable sum rates of the proposed AGC-LRA-MMSE linear
scheme.}
\label{fig:sum_rate_linear}
\end{figure}

In the following results we investigate the sum rates achieved by the proposed AGC-LRA-MMSE receiver by using both linear and SIC detection schemes. Fig. \ref{fig:sum_rate_linear} compares the achievable sum rates by the AGC-LRA-MMSE linear receiver and the sum rates achieved by the FR Standard MMSE receiver. In this result is possible to see that, the proposed
AGC-LRA-MMSE linear receiver achieves a sum rate similar to the sum rate achieved by the FR Standard MMSE receiver, even in a system whose signals are quantized
with only $6$ or $5$ bits.

\begin{figure}[H]
\centering
\includegraphics[scale=0.35]{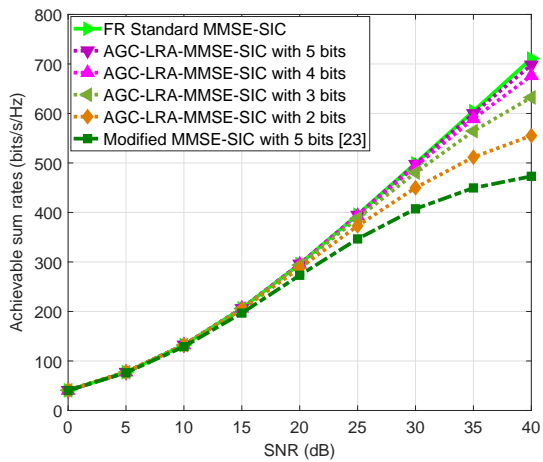}
\caption{Achievable sum rates of the proposed AGC-LRA-MMSE-SIC
scheme.}
\label{fig:sum_rate_sic}
\end{figure}

In Fig. \ref{fig:sum_rate_sic} we investigate achievable sum rates when the proposed
AGC-LRA-MMSE receiver is employed with SIC detection scheme. As expected, the AGC-LRA-MMSE-SIC scheme achieves a higher sum rate than that of the linear AGC-LRA-MMSE
receiver due to the SIC detection technique that improves the SINR of each
stream by the interference removal of the streams already detected. Similarly to the linear case, the sum rates achieved by the AGC-LRA-MMSE-SIC scheme in a system whose
signals are quantized with $5$ bits is close to the sum rates achieved by the FR Standard MMSE-SIC receiver with unquantized signals.

\section{Conclusions}

In this paper we have proposed the AGC-LRA-MMSE receiver design that jointly optimizes the AGCs that work in the RRHs and the receive filters that work in the CU for large-scale MU-MIMO systems in C-RANs with low-resolution quantized signals. The optimized AGC adjusts the dynamic range of the input signals inside the range of the quantizer in order to reduce the \textit{overload} and the \textit{granular} distortions. Their coefficients are calculated taking into account both the presence of the receive filter and the impact
of quantization, which implies in a more accurate AGC than existing techniques. The proposed design has been
incorporated into SIC detection scheme, resulting in substantial performance advantages over existing approaches. In particular, for QPSK modulation the
AGC-LRA-MMSE-SIC design can save up to $3$ dB in SNR in comparison to the best known approach \cite{roth} for the same BER performance, whereas the gains in achievable sum rate are up to about $45\%$ over the best known approach \cite{roth}. Furthermore, the sum rates and the BER achieved are
very close to those of unquantized systems for signals quantized with $5$ or $6$ bits. Therefore, the proposed AGC-LRA-MMSE-SIC design allows the use of low-resolution ADCs in large scale MU-MIMO systems with C-RANs that are important to improve the energy efficiency of
wireless systems and to compress signals, alleviating the capacity
bottleneck of the FH links. In particular, is important to mention that this paper presents the first design of an AGC designed and evaluated for large-scale MU-MIMO systems in C-RANs.

\appendix[Optimal AGC coefficients]

%
In this section we compute the derivatives of the cost function used
to obtain the optimum AGC matrices that were presented in Section
III. To compute each term of (\ref{eq:deriv_1}) we
consider the following property:
    \begin{eqnarray}
    \frac{\partial \tr[\mathbf{A}\diag(\mathbf{g})\mathbf{B}]}{\partial \mathbf{g}} = (\mathbf{A}^T\odot\mathbf{B})\mathbf{1},
    \end{eqnarray}
\noindent where $\mathbf{A}$ and $\mathbf{B}$ are complex matrices,
$\mathbf{g}$ is a vector with real coefficients and $\mathbf{1}$ is
a vector of ones. With this property we can take the derivative of
terms I, II, III, IV and V from (\ref{eq:deriv_1}). The derivatives
of the terms I and II are computed by
    \begin{eqnarray}
    \mathrm{I}&=&\frac{\partial \tr[\mathbf{R}_{xy}\diag(\mathbf{g}_l)\mathbf{W}_l^H]}{\partial \mathbf{g}_l} = [(\mathbf{R}_{xy}^T\odot\mathbf{W}_l^H)\mathbf{1}], \nonumber \\
    \mathrm{II}&=&\frac{\partial \tr[\mathbf{W}_l\diag(\mathbf{g})\mathbf{R}_{xy}^H]}{\partial \mathbf{g}_l} = [(\mathbf{R}_{xy}^H\odot\mathbf{W}_l^T)\mathbf{1}].
    \end{eqnarray}

    To compute the derivative of term III we apply the chain rule
    \begin{eqnarray}
    \mathrm{III}=\underbrace{\frac{\partial \tr[\mathbf{W}_l\diag(\mathbf{g}_l)\mathbf{A}]}{\partial \mathbf{g}_l}}_{\mathrm{III.1}}+\underbrace{\frac{\partial \tr[\mathbf{B}\diag(\mathbf{g}_l)\mathbf{W}_l^H]}{\partial \mathbf{g}_l}}_{\mathrm{III.2}},
    \end{eqnarray}
    \noindent where $\mathbf{A}=\mathbf{R}_{y_ly_l}\diag(\mathbf{g}_l)\mathbf{W}_l^H$ and $\mathbf{B}=\mathbf{W}_l\diag(\mathbf{g}_l)\mathbf{R}_{y_ly_l}$. The derivatives of terms III.1 and III.2 are computed by
    \begin{eqnarray}
    \mathrm{III.1} = [(\mathbf{W}_l^T\odot[\mathbf{R}_{yy}\diag(\mathbf{g}_l)\mathbf{W}_l])\mathbf{1}], \\
    \mathrm{III.2}= [(\mathbf{R}_{yy}^T\diag(\mathbf{g}_l)\mathbf{W}_l^T)\odot\mathbf{W}_l^H)\mathbf{1}].
    \end{eqnarray}

    Then
    \begin{eqnarray}
    \mathrm{III}= &&[(\mathbf{W}_l^T\odot(\mathbf{R}_{yy}\diag(\mathbf{g}_l)\mathbf{W}_l^H))\mathbf{1}]+\nonumber \\
    &&[((\mathbf{R}_{yy}^T\diag(\mathbf{g}_l)\mathbf{W}_l^T)\odot\mathbf{W}_l^H)\mathbf{1}].
    \end{eqnarray}

    The derivatives of terms IV and V are obtained by {\footnotesize
   \hspace{-1em} \begin{eqnarray}
    \mathrm{IV}&=&\frac{\partial \tr[\mathbf{W}_l\diag(\mathbf{g}_l)\mathbf{R}_{yq}\mathbf{W}_l^H]}{\mathbf{g}_l}= [(\mathbf{W}_l^T\odot[\mathbf{R}_{yq}\mathbf{W}^H_l])\mathbf{1}], \\
   \mathrm{V}&=&\frac{\partial \tr[\mathbf{W}_l\mathbf{R}_{yq}^H\diag(\mathbf{g}_l)\mathbf{W}_l^H]}{\partial \mathbf{g}_l}=[((\mathbf{R}_{yq}^*\mathbf{W}_l^T)\odot\mathbf{W}_l^H)\mathbf{1}].
    \end{eqnarray}}

Substituting these results in (\ref{eq:deriv_1}) and equating the
derivatives to zero we obtain
    \begin{align}
     [(&\mathbf{W}_l^T\odot(\mathbf{R}_{yy}\diag(\mathbf{g}_l)\mathbf{W}_l^H))+ \nonumber \\
     & \hspace*{0.5cm}((\mathbf{R}_{yy}^T\diag(\mathbf{g}_l)\mathbf{W}_l^T)\odot\mathbf{W}_l^H)]\mathbf{1}
     \nonumber \\&=\frac{1}{\alpha}([(\mathbf{R}_{xy}^T\odot\mathbf{W}_l^H)\mathbf{1}]+[(\mathbf{R}_{xy}^H\odot\mathbf{W}_l^T)\mathbf{1}]\nonumber \\&-[(\mathbf{W}_l^T\odot(\mathbf{R}_{yq}\mathbf{W}_l^H))\mathbf{1}]-[((\mathbf{R}_{yq}^*\mathbf{W}_l^T)\odot\mathbf{W}_l^H)\mathbf{1}]).\label{eq:123}
    \end{align}

To compute $\mathbf{g}_l$ we write the first and second terms of the
equation with the index notation, manipulate the terms and then we
return to the matrix notation. We can write the first and the second terms as
     \begin{align}
     [(\mathbf{W}_l^T\odot(\mathbf{R}_{y_ly_l}\diag(\mathbf{g}_l)\mathbf{W}_l^H)\mathbf{1}]&= \nonumber \\
     \sum_{j=1}^{KN_T}\sum_{a=1}^{N_R}& W_{l_{ji}}R_{y_ly_l{ia}}g_{l_a}W_{l_{aj}}^H, \\
     [(\mathbf{W}_l^H\odot(\mathbf{R}_{y_ly_l}^T\diag(\mathbf{g}_l)\mathbf{W}_l^T)\mathbf{1}]&= \nonumber \\
     \sum_{j=1}^{KN_T}\sum_{a=1}^{N_R} & W^H_{l_{ij}}R_{y_ly_l{ai}}g_{l_{a}}W_{l_{ja}}.
     \end{align}

With some manipulations we can isolate the vector $\mathbf{g}_l$
     \begin{align}
     &[(\mathbf{W}_l^T\odot(\mathbf{R}_{y_ly_l}\diag(\mathbf{g}_l)\mathbf{W}_l^H)~~+ \nonumber \\
     &\hspace*{0.5cm}\mathbf{W}_l^H\odot(\mathbf{R}_{y_ly_l}^T\diag(\mathbf{g}_l)\mathbf{W}_l^T))\mathbf{1}] \nonumber \\
     &=\sum_{j=1}^{KN_T}\sum_{a=1}^{N_R}W_{l_{ji}}R_{y_ly_lia}g_{l_a}W_{l_{aj}}^H+\sum_{j=1}^{KN_T}\sum_{a=1}^{N_R}W^H_{l_{ij}}R_{y_ly_lai}g_{l_a}W_{ja} \nonumber \\
     &=\sum_{a=1}^{N_R} g_{l_a}([(\mathbf{W}^T_l\mathbf{W}^*_l) \odot \mathbf{R}_{y_ly_l}+(\mathbf{W}^H_l\mathbf{W}_l) \odot \mathbf{R}_{y_ly_l}^T]_{ia}) \nonumber \\
     &=[(\mathbf{W}_l^T\mathbf{W}^*_l) \odot \mathbf{R}_{y_ly_l}+(\mathbf{W}_l^H\mathbf{W}_l) \odot \mathbf{R}_{y_ly_l}^T]\mathbf{g}_l. \label{eq:isolate}
     \end{align}

Substituting (\ref{eq:isolate}) in (\ref{eq:123}) and solving it with
respect to $\mathbf{g}_l$ we have
     \begin{align}
     \mathbf{g}_l&=[(\mathbf{W}_l^T\mathbf{W}_l^*) \odot \mathbf{R}_{yy}+(\mathbf{W}_l^H\mathbf{W}_l) \odot \mathbf{R}_{yy}^T]^{-1} \nonumber \\
     &\cdot\frac{2}{\alpha}(Re([(\mathbf{R}_{xy}^T\odot\mathbf{W}_l^H)\mathbf{1}])
     - Re([(\mathbf{W}_l^T\odot(\mathbf{R}_{yq}\mathbf{W}_l^H))\mathbf{1}])), \label{eq:optimumg}
     \nonumber
     \end{align}
\noindent and the optimum AGC matrix can be written as
$\mathbf{G}_l=\diag(\mathbf{g}_l)$.


%
%

\end{document}